# Fluid–Structure Interaction and Underwater Hydrostatic Implosion of Thin-Walled Metallic Cylinders in Semi-Confined Conditions

**Bolaji Oladipo[1], Helio Matos[1], Arun Shukla[1], and Sumanta Das[1,2*]**

[1]Department of Mechanical, Industrial and Systems Engineering, University of Rhode Island, Kingston, RI, 02881, United States

[2]Department of Civil and Environmental Engineering, University of Rhode Island, Kingston, RI, 02881, United States

[*]Corresponding author: S. Das (sumanta_das@uri.edu)

## Abstract

This study performs a comprehensive numerical investigation into the dynamic behavior and fluid–structure interactions (FSI) of metallic cylinders undergoing hydrostatic collapse in semi-confined fluid environments using a structured Arbitrary Eulerian–Lagrangian (ALE) formulation in LS-DYNA. The numerical model reproduces the experimentally measured collapse pressure of 3.69 MPa and predicts the first water hammer peak with a 1.01% error, confirming high predictive fidelity. Following validation, the study explores the influence of material type (aluminum versus titanium), cylinder slenderness ratio (L/D), and confinement diameter on collapse behavior, pressure evolution, and fluid motion. Titanium cylinders, producing peak water hammer pressures exceeding 70 MPa, consistently exhibited sharper collapses, higher water hammer pressures, and greater kinetic and strain energy accumulations compared to aluminum, due to their higher stiffness and yield strength. Lower L/D ratios induced more abrupt collapses, while higher L/D ratios facilitated axisymmetric deformation with more gradual energy dissipation. Confinement diameter further modulated these trends, with larger fluid volumes enabling intensified jet formation, and elevated fluid velocity magnitudes. The simulation results offer mechanistic insights into the coupling between structure and fluid, revealing that the interplay of geometry, material stiffness, and boundary confinement governs collapse-induced energy transfer. Full-field FSI provided deeper insight into collapse dynamics by capturing critical features, such as radial jetting, peak fluid velocities, and internal cavitation, not evident in pressure-time histories alone. These findings provide quantitative guidance for the design and safety assessment of subsea pressure housings, marine pipelines, and confined underwater structural systems exposed to extreme hydrostatic loading.

## 1. INTRODUCTION

Submerged closed-cylinder structures can undergo catastrophic collapse when subjected to external hydrostatic pressure beyond their structural limits, a phenomenon widely recognized as implosion (Matos et al., 2024). This dynamic event is characterized by the rapid inward failure of the cylinders, resulting in the sudden release of stored potential energy as a high-amplitude pressure wave that propagates through the surrounding fluid medium (Grantham-Coogan et al., 2024). The implosion process has garnered significant interest in the fields of underwater wave propagation (Turner, 2007) and structural dynamics (Zheng et al., 2023) for applications such as underwater shock testing (Diwan et al., 2012; Oladipo et al., 2024) and structural resilience evaluation (Qu et al., 2017). The rapid collapse of a submerged cylinder under hydrostatic pressure releases a high-energy pressure wave that can inflict damage on adjacent components (Sun et al., 2022). The energy released during the hydrostatic implosion of a thin-walled metallic cylinder is not confined to the imploding structure itself, but it propagates into the surrounding environment (Chu et al., 2022). Before collapse, elastic strain energy stored in the cylinder walls is rapidly converted into kinetic energy as the walls accelerate inward, abruptly compressing the surrounding water body (Gupta et al., 2014a). This rapid displacement generates intense implosion shockwaves, characterized by sharp pressure gradients and peak overpressures that can exceed ambient hydrostatic pressure by several fold (Tilton and Shukla, 2025). These transient pressure fields can interact with neighboring structures, deforming or damaging them, and in clustered assemblies may even trigger secondary implosions (Zheng et al., 2024). A stark demonstration of this occurred on November 12, 2001, at the Super-Kamiokande neutrino observatory in Japan, where the failure of a single 50 cm photomultiplier tube (PMT) generated a shockwave that propagated through the water, causing a chain reaction that destroyed approximately 7,000 nearby PMTs (Suzuki, 2019).

Experimental studies on hydrostatic implosions of composite cylinders within confined, open-ended structures show that boundary conditions, such as enclosure stiffness and limited fluid egress, can induce water-hammer pulses at the closed end, significantly altering peak pressures and collapse kinetics compared to free-field scenarios (Matos et al., 2018; Reilly et al., 2025). These findings stress the need for a systematic and mechanistic investigation into fluid–structure interaction and energy transfer during implosions in semi-confined environments. Empirical studies consistently reveal that confinement has a strong influence on both collapse dynamics and the resulting pressure fields (Salazar and Shukla, 2020). However, the specific pathways of energy conversion, from elastic strain to kinetic energy to fluid impulse, remain insufficiently quantified under varied boundary conditions, presenting an excellent opportunity for further research to deepen our understanding of these complex interactions. Comparison of the dynamic pressure-time histories demonstrates that confinement fundamentally alters the implosion response. Confinement-induced amplification mechanisms have been reported in underwater shock and implosion studies, where boundary proximity increases peak overpressure and impulse through wave trapping and reflection effects (Gao et al., 2022). In contrast, unconstrained collapse permits radial energy dispersion and geometric attenuation of outgoing pressure waves, resulting in lower peak pressures and faster decay rates, consistent with classical underwater blast scaling behavior (Zhang et al., 2025). These differences align with previously reported observations that confinement modifies the fluid momentum redistribution and increases effective loading on adjacent structures. Semi-confined implosions involving thin-walled cylinders encased in rigid enclosures (Oladipo et al., 2026) exhibit markedly different collapse behaviors compared to free-field events (Gupta et al., 2015). Experiments with aluminum cylinders housed in pressure vessels, featuring open-end flow areas, have shown confinement can delay collapse initiation, alter buckling modes, and produce amplified water-hammer spikes upon fluid deceleration (Shukla et al., 2018). These deviations suggest that confinement geometry directly influences fluid acceleration and reflection patterns, thereby modifying peak pressures and collapse timelines. Extensive research has elucidated the formation and propagation of implosion-generated pressure waves, as well as their dependence on cylinder geometry. While prior studies have significantly advanced our understanding of pressure wave formation and propagation, especially as influenced by cylinder geometry, a substantial knowledge gap persists regarding the role of single open-ended confinement. This is particularly true when material properties and structural geometries of surrounding environments vary, complicating predictions of collapse outcomes and pressure signatures.

While experimental investigations provide critical insights and ground-truth validation, they are often constrained by high costs, safety concerns, limited diagnostic access, and difficulties in precisely controlling boundary conditions. These challenges make it difficult to systematically explore a wide range of geometries and confinement scenarios. Given the destructive potential and complex physics of implosion events, high-fidelity numerical models offer a complementary and, in many cases, more flexible approach for probing detailed fluid–structure interactions, enabling researchers to isolate specific parameters, replicate complex environments, and predict implosion behavior with greater control and resolution. Mostly, the predictive models developed so far have been based on simplified analytical solutions (Cor and Miller, 2010; Von Mises and Windenburg, 1933), which gained popularity due to their ease of implementation, low computational cost, and ability to provide quick, approximate insights into structural response. For instance, Farhat et. al. (Farhat et al., 2013) developed a theoretical model to predict dynamic buckling and FSI in submerged cylindrical cylinders, elucidating how variations in cylinder stiffness, fluid properties, and hydrostatic pressure influence collapse initiation and progression. However, these analytical models often rely on idealized boundary conditions and overlook complex fluid–structure interactions, making them inadequate for capturing the nonlinear collapse mechanisms, localized deformations, and transient pressure fields observed in actual implosion events. Computational studies of fluid–structure interaction (FSI) during underwater structural collapse remain limited, largely due to the high computational cost, complexities in modeling the coupled fluid and structural domains, and challenges in experimental validation. Nonetheless, a few specific numerical studies have employed advanced simulation frameworks (Oladipo et al., 2023; Oladipo and Das, 2026; Villada et al., 2023), most notably DYSMAS (Chambers et al., 1998) and LS-DYNA (Livermore Software Technology Corporation) (Turner and Ambrico, 2012), to investigate these interactions under extreme hydrostatic loading. For example, Turner et. al. (Turner and Ambrico, 2012) utilized DYSMAS (Chambers et al., 1998) to simulate the hydrostatic implosion of 6061-T6 aluminum cylinders, achieving excellent agreement with experimental data and demonstrating coupled fluid and structural dynamics in replicating collapse pressure–time profiles. In a separate study, Wei et al. (Wei et al., 2020) applied LS-DYNA in supercomputer-based simulations to reconstruct the implosion sequence of the ARA San Juan S-42 submarine, capturing nonlinear collapse behavior, fragment dispersal, and shock radiation comparable to observational evidence. Large-scale numerical investigations, such as the supercomputer-based simulations by Wei et al., demonstrated that LS-DYNA can reproduce complex submarine implosion sequences, including nonlinear collapse progression and shock radiation patterns broadly consistent with reported observations. However, the validation in these studies was largely qualitative and phenomenological in nature. Quantitative error margins for key metrics such as peak pressure, collapse timing, and mesh sensitivity were not systematically reported. Moreover, the reconstruction of the ARA San Juan implosion involved a large-scale and geometrically complex submarine configuration under deep-water conditions without direct experimental replication. As a result, predictive accuracy was inferred indirectly from consistency with reported acoustic signatures and debris observations rather than from controlled laboratory benchmarks. In addition, the open-ocean conditions considered in these studies differ fundamentally from semi-confined geometries, where boundary proximity and wave reflections significantly influence pressure amplification and fluid–structure interaction dynamics. These limitations highlight the need for controlled configurations in which collapse pressure, overpressure magnitude, wave propagation, and deformation modes can be rigorously compared with laboratory measurements. The present study addresses this gap by implementing a structured ALE fluid–structure interaction framework for thin-walled metallic cylinders in semi-confined environments, enabling direct experimental–numerical comparison and systematic evaluation of the effects of material stiffness, cylinder slenderness ratio (L/D), and confinement diameter on implosion dynamics.

This study advances the understanding of semi-confined underwater implosions through a combined experimental and numerical approach. Simulation fidelity is ensured by validating numerical predictions against an experimentally established test protocol that closely replicates the simulated conditions. The experimental setup mirrors the numerical model's material properties, geometry, boundary conditions, and hydrostatic pressure profile. Key collapse metrics, such as initiation pressure, wall deformation, collapse timing, and transient pressure waveforms, are directly compared. To fully capture the underlying mechanisms, this study integrates three coordinated components: (1) Fluid–structure interaction (FSI) simulations using Arbitrary Eulerian–Lagrangian (ALE) methods (Luo et al., 2025) within fully coupled LS-DYNA models, which capture transient structural deformation, fluid acceleration, and wave reflections during collapse; (2) A parametric investigation of cylinder geometries and material properties, examining how variations in length-to-diameter ratio, elastic modulus, and yield strength influence collapse thresholds, buckling patterns, energy transfer, and pressure pulse characteristics; and (3) A systematic study of confinement effects, in which high-fidelity implosion simulations are conducted across a range of confinement geometries, including variations in enclosure stiffness and open-end area. Through this integrated approach, the study systematically explores the complex interplay between geometric parameters, material behavior, and confinement conditions in driving underwater structural collapse. By systematically isolating the effects of geometry, material properties, and boundary constraints, this study provides new insights into the mechanics of implosion under semi-confined conditions, bridging critical gaps between simplified analytical models, experimental observations, and high-fidelity numerical predictions. The integrated framework developed here not only enhances predictive capabilities for implosion-driven pressure events but also lays the groundwork for designing safer, more resilient underwater systems in applications where structural collapse and shock transmission pose significant risk.

## 2. NUMERICAL SIMULATION METHOD

LS-DYNA was selected for this study due to its advanced capabilities in simulating complex fluid–structure interaction (FSI) under extreme dynamic loading conditions. As a robust explicit finite element solver, LS-DYNA effectively handles large deformations, nonlinear material responses, and highly transient fluid–structure coupling, making it well-suited for capturing the intricate collapse and fluid dynamics involved in these scenarios. LS-DYNA solver incorporates advanced contact algorithms, large strain plasticity models, and robust equation of state formulations, enabling accurate treatment of geometric nonlinearity, material yielding, and shock-compressible fluid response within a unified framework (Haghgoo et al., 2022, 2021; Hallquist, 1994; Oladipo et al., 2026). The following subsections detail the geometry generation for both the metallic cylinders and surrounding fluid domains, the FSI coupling strategy, boundary conditions, and the relevant geometric features and material properties used in the simulations.

### 2.1 Geometry Generation and Boundary Conditions

The water domain is modeled using LS-DYNA's Arbitrary Lagrangian–Eulerian (ALE) framework (Wei et al., 2020), a robust approach for resolving strong fluid–structure interaction phenomena. The Structured ALE (S-ALE) solver in LS-DYNA (Hao Chen, 2016), introduced in 2015, is optimized for structured-mesh applications that require regular, box-shaped mesh configurations. It features a built-in automated mesh generator that constructs rectilinear hexahedral meshes directly from user-defined nodal control points, eliminating the need for pre-generated mesh input and reducing simulation setup time and file size. S-ALE allows for variable element sizes via progressive spacing, enabling mesh refinement in critical regions while keeping a coarser mesh elsewhere to conserve computational resources. Benefiting from structured connectivity, S-ALE also achieves improved efficiency, reduced memory usage, and enhanced parallel performance (Hao Chen, 2016). The solver enables multi-material modeling by allowing multiple materials to coexist within a single Eulerian mesh, which facilitates high-fidelity simulation of complex material interactions coupled with fluid dynamics.

By assigning each material phase to a distinct *ALE_MULTI-MATERIAL_GROUP, the solver maintains precise interface reconstruction, ensuring that volume fractions of disparate materials, such as water, air, and structural solids, are resolved within each cell. This approach is particularly well-suited for simulating advanced underwater structures under implosive loading conditions, where accurate resolution of interface dynamics and transient wave propagation is critical (Wang et al., 2025).

The water domain utilizes the *MAT_NULL material model, a straightforward formulation for fluid-like media having a typical density of 0.001 g/mm$^3$ and sound speed of 1490 mm/ms, which is employed to replicate a realistic acoustic response. Mesh resolution and time-step size are calibrated according to the Courant–Friedrichs–Lewy (CFL) criterion, striking a balance between computational precision and efficiency. The time-step is constrained by the condition that numerical wave propagation must not exceed one element length per step, i.e., $\Delta t \leq CFL * (\ell_min / c)$, where $\ell_min$ is the smallest mesh dimension and c is the local sound speed. Mesh elements are refined in regions expecting steep gradients (e.g., near cylinder walls or boundaries) while coarser elsewhere enables resource management (Ikeda et al., 2013). Fig. 1 illustrates the discretized water domain employed to simulate FSI interaction during hydrostatic implosion in a semi-confined environment. The domain is constructed as a structured, progressive mesh: mesh density is highest in the vicinity of the cylinder, where steep pressure gradients and rapid deformation occur, and progressively coarsens outward to conserve computational resources while preserving global accuracy. The cuboidal fluid domain of size 300 × 300 × 1800 mm$^3$ offers sufficient clearance around the structure to mitigate artificial wave reflections when used in conjunction with non-reflecting boundary conditions. This mesh strategy ensures high-fidelity capture of implosion shockwaves and fluid momentum transfer near the structure, while minimizing element count and CPU time in less critical regions. To ensure numerical accuracy and independence of discretization parameters, a mesh and time-step sensitivity study was performed. Five progressively refined structured meshes were evaluated. The baseline configuration employed an average element size of 2 mm near the implodable cylinder and progressively coarsened elements toward the far-field boundaries. Other evaluated baseline average element size were 4 mm, 3 mm, 1 mm and 0.5 mm. Key response metrics including collapse time, first water hammer peak pressure, and wave arrival time were monitored. Differences between the baseline and finest mesh remained below 2.5% for peak pressure and below 2% for collapse timing, confirming spatial convergence of the solution.

The computational domain dimensions of 300 × 300 × 1800 mm were selected to prevent artificial boundary reflections from influencing the implosion response within the analysis time window. A domain extension study, in which the axial length was increased by 20%, resulted in less than 1 percent variation in peak overpressure and collapse time, confirming domain adequacy.

Time integration stability and accuracy were controlled using the Courant–Friedrichs–Lewy condition with automatic time-step adjustment. Simulations were conducted with CFL numbers of 0.6, 0.4, and 0.3 to assess temporal sensitivity. Variations in predicted peak pressure and collapse timing remained below 1.5% across these cases, demonstrating temporal convergence. These results confirm that the reported implosion dynamics are independent of mesh resolution, domain size, and time-step selection within acceptable numerical tolerance.

The *BOUNDARY_NON_REFLECTING boundary condition is applied to all outer surfaces of the water domain to emulate an unbounded marine environment. This formulation minimizes artificial wave reflections at the mesh boundaries by matching the local fluid impedance, ensuring that outgoing pressure waves are absorbed rather than reflected into the domain. Hydrostatic implosion is induced by incrementally increasing pressure via dynamic loading applied across all six non-reflecting boundary faces until the cylinder reaches its critical collapse pressure. Under this uniform loading condition, no propagating pressure fronts interact with the domain boundaries, and therefore the non-reflecting boundaries do not influence the pre-collapse stress field. Following structural instability, collapse generates strong outward-propagating pressure waves and water hammer pulses. The effectiveness of the non-reflecting boundaries during this transient phase was evaluated through comparative simulations. In a reference case, fixed pressure boundaries were applied instead of non-reflecting boundaries, resulting in amplified secondary oscillations and measurable late-time pressure distortion due to artificial reflections. In contrast, the non-reflecting boundary configuration showed stable attenuation of outgoing waves, with less than 3% variation in peak pressure when compared to an extended-domain simulation.

Additionally, the domain dimensions were selected such that the travel time of reflected waves from the outer boundaries exceeds the time window of the primary implosion and first water hammer event. This ensures that boundary reflections do not contaminate the reported collapse pressure, peak overpressure, or early-time wave propagation characteristics. These assessments confirm the suitability and effectiveness of the non-reflecting boundary treatment for the present semi-confined hydrostatic implosion simulations.

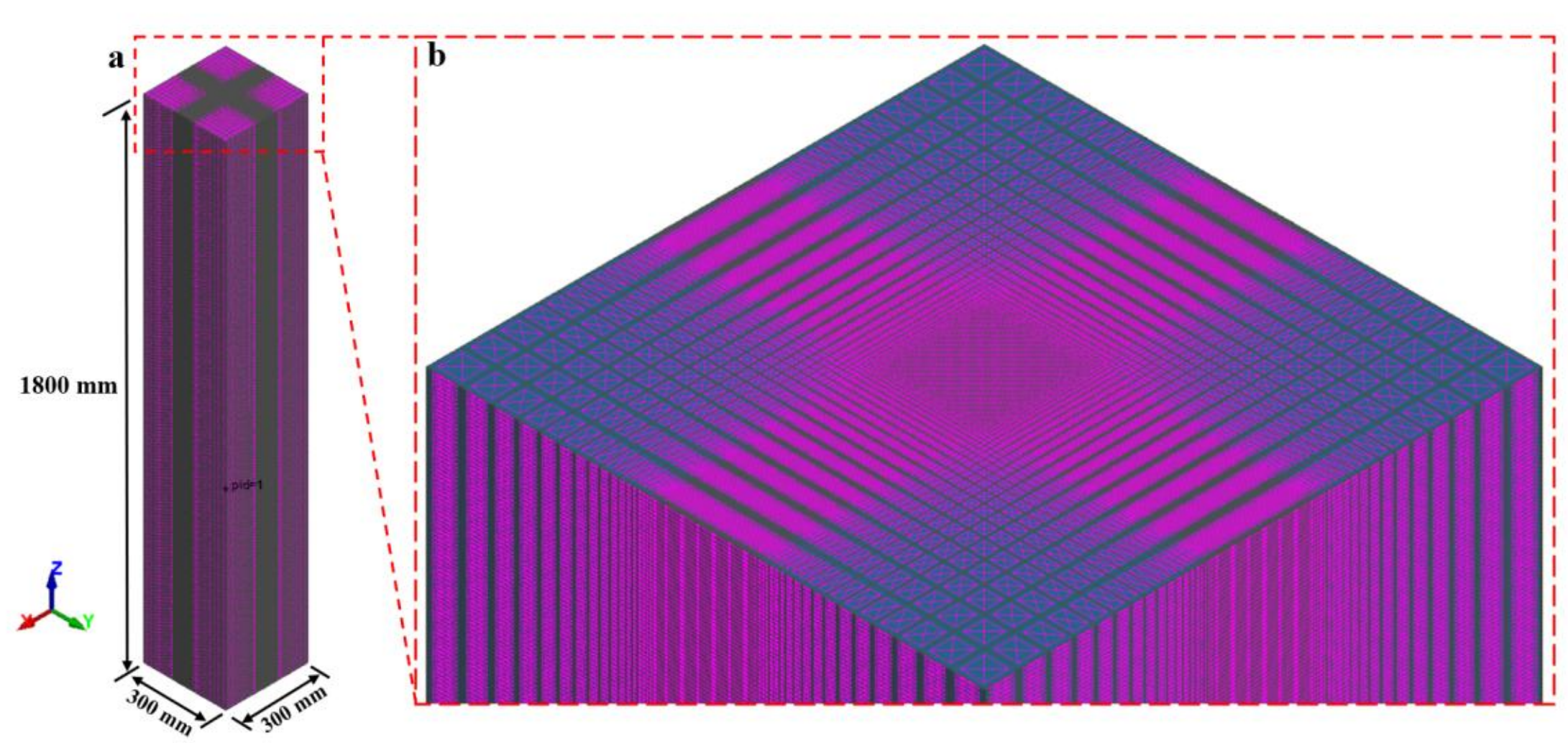

Fig. 1: Discretized water domain using a structured, (a) progressively refined mesh within a $300 \times 300 \times 1800$ mm³ cuboid, optimized for accurate fluid–structure interaction modeling during hydrostatic implosion in a semi-confined environment (b) zoomed view of the top surface of the discretized water domain showing progressively refined mesh

The air in the tube is created using the *INITIAL_VOLUME_FRACTION_GEOMETRY keyword (Wei et al., 2020). The *INITIAL_VOLUME_FRACTION_GEOMETRY keyword in LS-DYNA streamlines multi-material ALE fluid domain initialization by allowing users to specify geometrical primitives, such as cylinders, boxes, or spheres, to automatically fill the Eulerian mesh with designated materials (e.g., air). This feature eliminates the need for manual mesh partitioning, significantly reducing setup complexity and potential for user error. By assigning material volume fractions based on predefined shapes, the method ensures accurate representation of fluid–structure boundaries and enhances computational efficiency, an efficiency that is further improved through reduced memory and disk usage (Aquelet et al., 2005). Air is modeled with a density of $1.3 \times 10^{-6}$ g/mm³ using the *MAT_NULL material model at approximating sea-level conditions.

Fig. 2 illustrates the LS-DYNA finite-element model of an aluminum cylindrical implodable tube in a one-end-open, one-end-sealed confining cylinder to study hydrostatic implosion in a semi-confined setting. The confining cylinder spans 1270 mm in length, with an outer diameter of 178 mm and a wall thickness of 25.4 mm. Inside, the implodable cylinder measures 302 mm in length and 25.4 mm in outer diameter, resulting in a slender geometry with an L/D ratio of approximately 11.9. The cylinder's ends are rigidly capped to emulate boundary conditions commonly encountered in submerged or constrained systems.

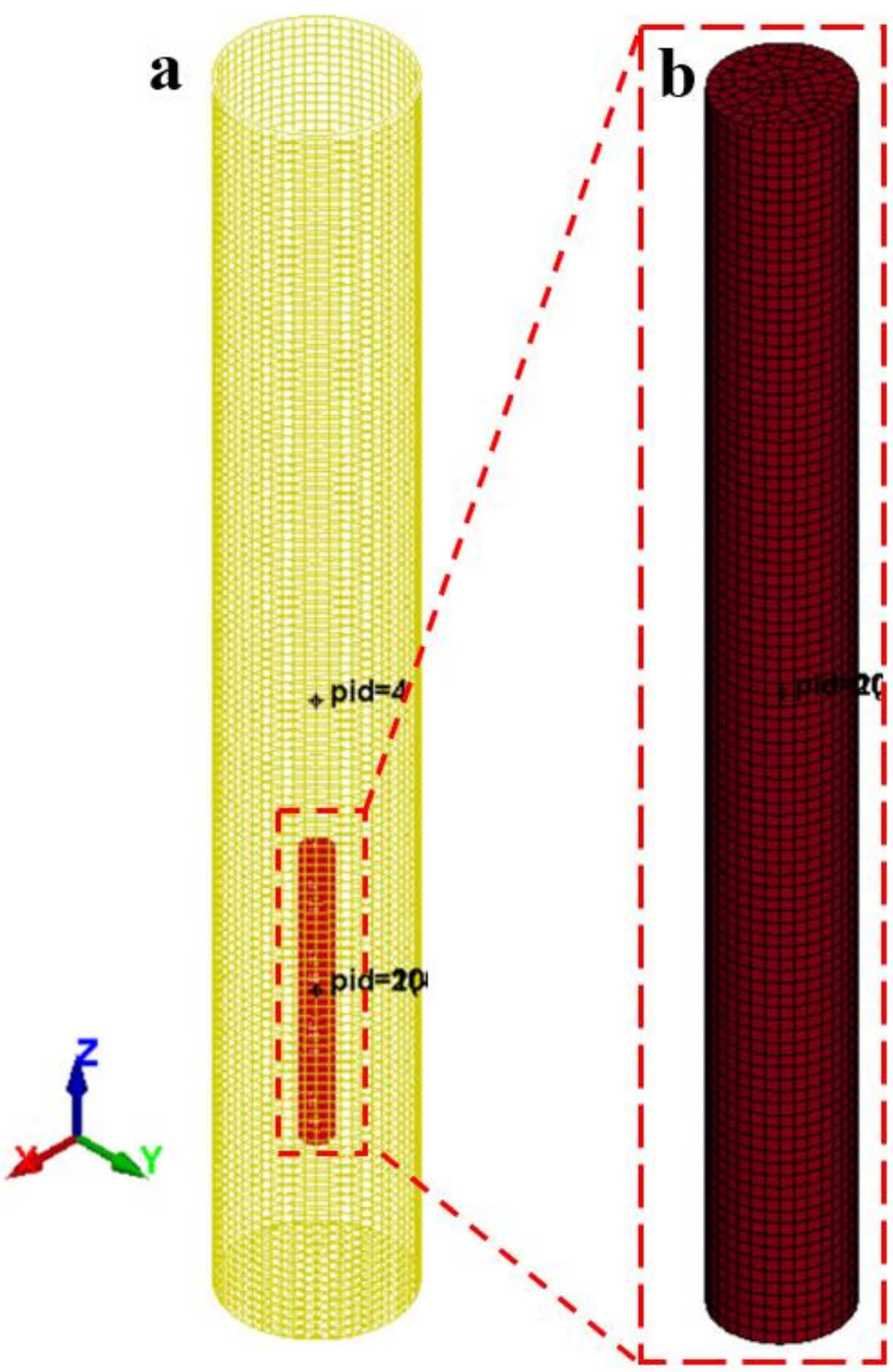

Fig 2: LS Dyna model of (a) confining cylinder structure – aluminum implodable assembly (b) aluminum cylinder with a dimension of 302 mm in length, a diameter of 25.4 mm, and a thickness of 0.87 mm.

**2.2. Governing Equations of State**

This study implements the Grüneisen Equation of State (*EOS_GRUNEISEN) for water and the Linear Polynomial EOS (*EOS_LINEAR_POLYNOMIAL) for air to model their thermodynamic responses under dynamic loading (Arienti et al., 2004).

### 2.2.1. Gruneisen Equation of State for Water

The Grüneisen EOS for water captures pressure as a function of density and specific internal energy, effectively representing compressibility and shockwave behavior in high-pressure environments (Wang et al., 2024). The Mie–Grüneisen equation of state (EOS) is particularly effective for capturing the behavior of materials under high-pressure and shock-loading conditions, such as during hydrostatic implosions. In its most commonly used form, this EOS relates pressure to specific volume and internal energy by equation 1 below (Löhner et al., 2023):

$$p = p_{\mathrm{H}} + \Gamma\rho(E - E_{\mathrm{H}}) \tag{1}$$

Where $p_{\mathrm{H}}$ is the Hugoniot pressure, $\rho$ is the density, $p$ is the pressure, $\Gamma$ is the Grüneisen coefficient, $E_{\mathrm{H}}$ is the Hugoniot internal energy, and $E$ is the specific internal energy. The Grüneisen (EOS) effectively captures shock-induced energy variations, making it well-suited for modeling water under implosive conditions. By relating pressure to specific volume and internal energy, the Grüneisen EOS adeptly represents the behavior of compressible fluids experiencing large-volume changes and dynamic loading. Therefore becomes:

$$p = \frac{\rho_0 C^2 \mu\left[1+\left(1-\frac{\gamma_0}{2}\right)\mu-\frac{a}{2}\mu^2\right]}{\left(\left[1-(S_1-1)\mu-\frac{S_2\mu^2}{\mu+1}-\frac{S_3\mu^3}{(\mu+1)^2}\right]^2+(\gamma_0+a\mu)E\right)} \tag{2}$$

Which further becomes:

$$p = \rho_0 C^2\mu + (\gamma_0 + a\mu)E \tag{3}$$

Where $C$ is the velocity intercept curve, $\rho_0$ is the density of reference, $\gamma_0$ is the coefficient of Gruneisen gamma, $S_1, S_2$, and $S_3$ are coefficients that define the curve slope, and $\mu$ is $\rho/\rho_0 - 1$, with $E$is the internal energy and $a$ represents the first-order volume correction.

Eq. 3 represents a reduced form of the full Mie–Grüneisen formulation presented in Eq. 2. This simplification is introduced under the assumption of a linear shock velocity–particle velocity relationship for water within the moderate pressure regime relevant to the present hydrostatic implosion simulations. In the implemented model, the higher-order Hugoniot slope coefficients $S_2$ and $S_3$ were set to zero, and the first-order volume correction parameter a was taken as zero. These assumptions eliminate higher-order nonlinear compressibility contributions that become significant only at extreme shock pressures substantially exceeding those observed in this study.

The pressure levels generated during collapse and subsequent water hammer events remain within the tens of megapascal range, where the linear Hugoniot approximation provides accurate representation of water compressibility. Cavitation effects are represented through tensile cutoff behavior in the fluid response rather than through modification of the equation of state. No sustained vapor phase evolution or phase transition was observed in the experimental validation cases; therefore, multiphase thermodynamic modeling was not required.

Under these loading conditions, the simplified formulation retains the essential pressure–volume–energy coupling necessary to accurately capture wave propagation and shock response while avoiding unnecessary higher-order complexity. The close quantitative agreement between simulation and experiment, including collapse pressure and peak overpressure prediction, further confirms the adequacy of this reduced Mie–Grüneisen representation for the present semi-confined implosion configuration.

### 2.2.2 Polynomial Equation of State for Water

For air, the Linear Polynomial EOS approximates ideal gas behavior by defining pressure linearly with internal energy and density fluctuations, making it computationally efficient and well-suited for simulating compressible fluid dynamics.

$$P = C_0 + C_1\mu + C_2\mu^2 + C_3\mu^3 + (C_4 + C_5\mu + C_6\mu^2)E \tag{4}$$

Equation 4 conforms with the EOS gamma law, where $C_0 = C_1 = C_2 = C_3 = C_6 = 0$ and $C_4 = C_5 = \gamma - 1$, where $\gamma$ is the air's specific heat ratio.

The compressible response of water was modeled using the Mie–Grüneisen equation of state. To ensure full reproducibility and transparency, all EOS parameters employed in the simulations are explicitly reported in Table 1. The selected parameters correspond to standard compressible water properties widely adopted in underwater shock and implosion simulations and are consistent with validated LS-DYNA material libraries. These values accurately capture the acoustic wave speed, compressibility, and pressure–volume relationship within the pressure range encountered in the present hydrostatic implosion scenarios. The internal air domain was modeled using the Linear Polynomial equation of state with ideal gas behavior. The selected parameters are appropriate for the pressure range observed prior to structural collapse and are sufficient to capture compressibility effects without requiring phase change modeling.

**Table 1: Equation of State Parameters Used in the Numerical Simulations**

| Material | EOS Type | $\rho_0$ (g/mm³) | C (mm/µs) | $S_1$ | $S_2$ | $S_3$ | $\gamma_0$ | a | $\gamma$ |
|---|---|---|---|---|---|---|---|---|---|
| Water | Mie–Grüneisen | 0.001 | 1.49 | 1.79 | 0.0 | 0.0 | 0.11 | 0.0 | - |
| Air | Linear Polynomial (Ideal Gas) | $1.3 \times 10^{-6}$ | - | - | - | - | - | - | 1.4 |

## 2.3. Modeling Assumptions and Limitations

The present numerical framework incorporates several modeling assumptions that define the scope of applicability of the results. The compressible response of water was modeled using a simplified Mie–Grüneisen equation of state. In the pressure regime relevant to the present hydrostatic implosion scenarios, higher-order Hugoniot slope terms and volume correction parameters were neglected, which is consistent with the moderate pressure range investigated in this study. Multiphase effects, phase transition, and detailed cavitation bubble dynamics were not explicitly modeled; instead, cavitation onset was represented through a tensile cutoff formulation in the fluid response. Structural fracture and fragmentation of the collapsing cylinders were also not included, as the primary objective of the study was to investigate collapse dynamics, pressure amplification, and fluid–structure interaction during the implosion event rather than the subsequent debris evolution. Furthermore, the structured ALE formulation treats the fluid as a continuum medium and does not resolve microscale turbulence or viscosity-dominated flow effects. Finally, the confinement geometry used in the simulations represents an idealized configuration. Although geometric imperfections were incorporated based on measured parameters from the experimental specimens, these values cannot capture the full range of manufacturing variability that may occur in practical structures.

### 2.4. Geometric and Material Parameters

The geometric parameters of the cylinder and confinement, together with the mechanical properties of each alloy, are considered to assess structural behavior and deformation under varying load conditions. Table 2 presents the material properties incorporated into the numerical simulations for two metallic alloys, which are 6061-T6 aluminum and Ti-6Al-4V (TC4) titanium. Aluminum exhibits an elastic modulus of approximately 68.9 GPa and a density of 2.70 g/cm$^3$, while titanium shows significantly higher values at 114 GPa elastic modulus and 4.42g/cm$^3$ density. Both materials share a Poisson's ratio of 0.33. Ultimate tensile strength also differs, with aluminum rated at 310 MPa and Ti-6Al-4V reaching 903 MPa. The cylinder material is represented using the *MAT_PLASTIC_KINEMATIC (MAT_003) (Wei et al., 2020) material model in LS-DYNA, an isotropic elastoplastic constitutive law that effectively captures plastic deformation behavior.

Table 2: Material properties for numerical simulation

| | Aluminum T6-6061 alloy **(Hellier et al., 2017)** | Titanium TC4 alloy **(Ji et al., 2014; Wang et al., 2022)** |
|---|---|---|
| Density (g/cm$^3$) | 2.70 | 4.42 |
| Elastic modulus (GPa) | 68.9 | 114 |
| Ultimate tensile strength (MPa) | 310 | 903 |
| Poisson's ratio | 0.33 | 0.33 |

The numerical simulations employed cylinders with length-to-diameter (L/D) ratios of 2 and 5, selected (Ikeda et al., 2013), and maintained a constant wall thickness of 1 mm, classifying the cylinders as thin-walled (Khan and Kumar, 2025). Semi-confined scenarios were modeled using enclosure diameters of 150 mm and 250 mm to explore the effects of geometry and material on collapse behavior. This comprehensive parametric design allows the evaluation of stiffness, density, strength, aspect ratio, wall thickness, and confinement geometry jointly influence collapse initiation pressure, buckling mode progression, implosion timing, and pressure-wave emission.

## 3. RESULTS AND DISCUSSION

The results and discussion section begins by validating the numerical model against experimental data, ensuring accurate representation of collapse behavior under matched conditions. It then presents a detailed analysis of how material selection, specifically 6061-T6 aluminum versus Ti-6Al-4V titanium and geometrical attributes, such as the length-to-diameter (L/D) ratio of the implodable cylinders, influence implosion dynamics. Furthermore, it examines the impact of confining cylinder geometries, with particular attention to variations in open-end cross-sectional area, on deformation modes, energy absorption characteristics, and the resulting dynamic pressure responses.

### 3.1 Experimental Validation

#### *3.1.1 Experimental Setup*

The experimental validation of the numerical model was conducted using a 302 mm long aluminum cylinder (outer diameter 38.1 mm, wall thickness 0.87 mm), rigidly sealed at both ends to encapsulate air and prevent water ingress. This closed cylinder was submerged in a water-filled semi-confined enclosure measuring 1270 mm in length, with an outer diameter of 178 mm and a wall thickness of 25.4 mm. The configuration mirrored the numerical setup, enabling direct comparison of hydrostatic implosion behavior.

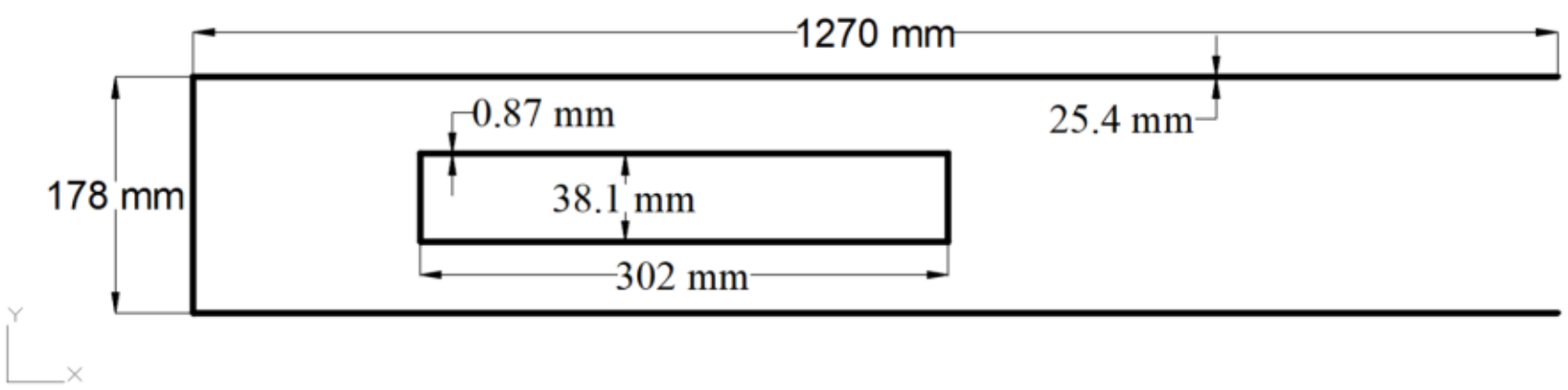

Fig. 3: 302 mm long aluminum cylinder (38.1 mm outer diameter, 0.87 mm wall thickness) placed inside a semi-confined enclosure measuring 1270 mm in length, 178 mm in outer diameter, and 25.4 mm in wall thickness.

Fluid pressure within the enclosure was increased incrementally until the cylinder reached its collapse pressure. High-speed imaging paired with synchronized transient pressure transducers recorded wall deformation, collapse timing, and pressure pulse evolution. This methodology is consistent with previous studies on implodable cylinders in confining geometries, facilitating accurate correlation between measured and simulated fluid–structure interaction phenomena

Implodable volumes were fabricated from commercially available 6061-T6 seamless aluminum tubing with an outer diameter of 38.1 mm and a nominal wall thickness of 0.870 mm. Each specimen was cut from a single 1.8 m length of extruded tubing. Initial geometric imperfections were quantified: cross-sectional ovality, defined as $\Delta_0 = (a_{max} - a_{min})/(a_{max} + a_{min})$, was found to be 0.01%; wall-thickness eccentricity, defined as $\Xi_0 = (h_{max} - h_{min})/(h_{max} + h_{min})$, measured 2.53% (E. John Finnemore and Edwin P. Maurer, 2023). To prevent fluid ingress, aluminum endcaps with O-ring seals were press-fitted at both ends of the cylinder [35]. Fig. 4 presents a detailed schematic of the experimental setup used to investigate the hydrostatic implosion of slender metal cylinders within a semi-confined environment. The test apparatus is situated within a 2.13 m-diameter underwater pressure vessel. Surrounding the cylinder, floodlights and a high-speed camera are oriented through the vessel's viewing port, mounted at a nominal 17° angle, to capture the implosion event with minimal optical distortion. A comprehensive description of the facility is provided in (Nayak et al., 2022). Centrally located inside a 1270 mm long, 178 mm internal diameter confining cylinder constructed from 6061-T6 aluminum with 25.4 mm wall thickness, is the implodable aluminum cylinder. One end of the cylinder is rigidly sealed, while the opposite is left open to the pressurizing water, thereby creating an asymmetric boundary condition. This implodable cylinder is capped at both ends to trap air and prevent water ingress. PCB-113B22 high-frequency dynamic pressure sensors are flush-mounted to the inner wall at multiple locations of the confining cylinder to capture rapid pressure transients. Once the pressure vessel is filled with water and pressurized (via a nitrogen gas inlet), incremental hydrostatic pressure is applied until the implodable cylinder catastrophically collapses. Data from the pressure sensor, recorded at a 2 MHz sampling rate, are synchronized with high-speed camera imaging to resolve collapse kinematics with high temporal fidelity. To investigate the evolution of water-hammer pressure waves induced by the implodable volume, experiments were conducted by placing the cylinder adjacent to the closed end of the confining cylinder (see Fig. 4c).

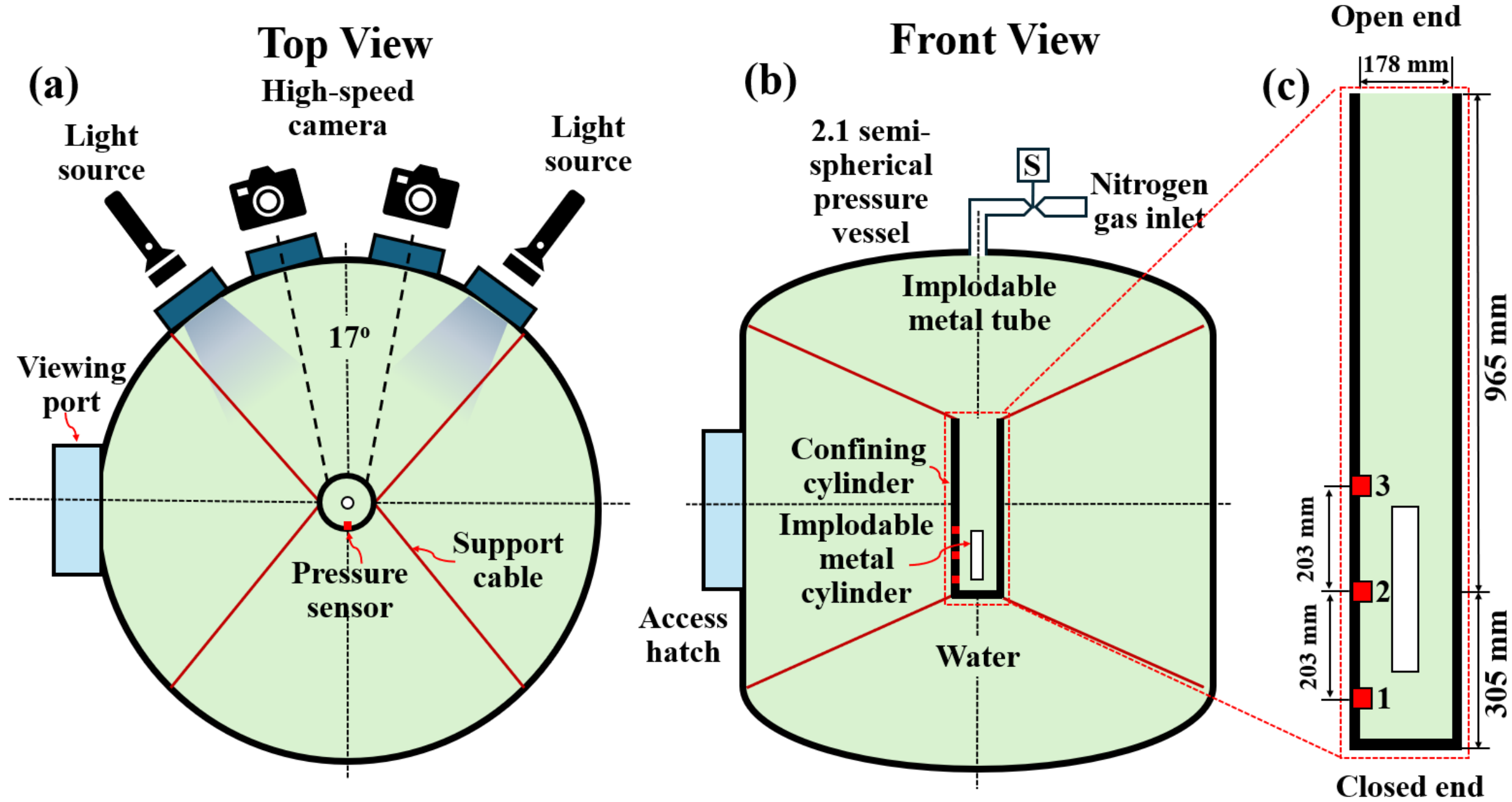

Fig. 4: Experimental configuration of the semi-confined implosion setup. (a) Top view inside the 2.13 m diameter underwater pressure vessel, showing the confining cylinder, dynamic pressure sensor (PCB 113B22), high-speed imaging system, and lighting arrangement. (b) Front view of the setup, illustrating the 1270 mm-long confining cylinder (178 mm internal diameter, 25.4 mm wall thickness) with one end sealed and the other open to pressurized water; the aluminum implodable cylinder is centrally positioned. (c) Sectional side view showing the placement of the specimen and pressure sensors: the bottom end is closed, the top open, and sensor 2 is located 305 mm from the confining cylinder base to the mid area of the implodable cylinder; other key enclosure dimensions and sensor positions are indicated.

The pressure measurements were obtained using calibrated high-frequency dynamic pressure transducers with a manufacturer-specified accuracy of ±1 percent of full-scale output. Given the pressure range recorded during implosion, this corresponds to an uncertainty substantially smaller than the measured peak overpressures. All pressure signals were acquired using a data acquisition system operating at a sampling rate of 2 MHz, providing a temporal resolution of 0.5 ms. The timing synchronization between the pressure transducers and the high-speed imaging system was verified prior to testing, with an estimated synchronization uncertainty below 0.1 ms. This timing error is negligible relative to the millisecond-scale collapse and water hammer events reported in this study.

Geometric imperfections of the implodable cylinders were quantified prior to testing to account for their influence on collapse behavior. The ovality parameter and wall eccentricity parameter were measured using precision dimensional inspection techniques and incorporated into the numerical model to maintain consistency between experimental and simulated configurations. Repeated tests under identical loading conditions showed minimal variation in collapse pressure, indicating good experimental repeatability. These quantified uncertainties provide a clear measurement confidence framework and support the reported agreement between experimental observations and numerical predictions.

#### *3.1.2 Comparison Between Experimental and Numerical Results*

Fig. 5 provides a comparative view of dynamic pressure-time responses during hydrostatic implosion in a semi-confined cylinder, showing experimental results in Fig. 5(a) and numerical simulation outputs in Fig. 5(b). Both datasets capture the key stages of the event: collapse initiation, cavitation bubble formation, and a successive water-hammer pulse. In both the experimental and numerical cases, implosion is initiated at a critical pressure of approximately 3.59 MPa. This triggers a sudden structural collapse accompanied by rapid inward fluid acceleration and cavitation, evidenced by a sharp pressure drop below critical collapse pressure in both figures. The formation and collapse of a cavitation bubble are clearly visible in each pressure trace. Both the experiment and simulation exhibited a Mode 2 buckling pattern, characterized by two circumferential lobes along the length of the cylinder.

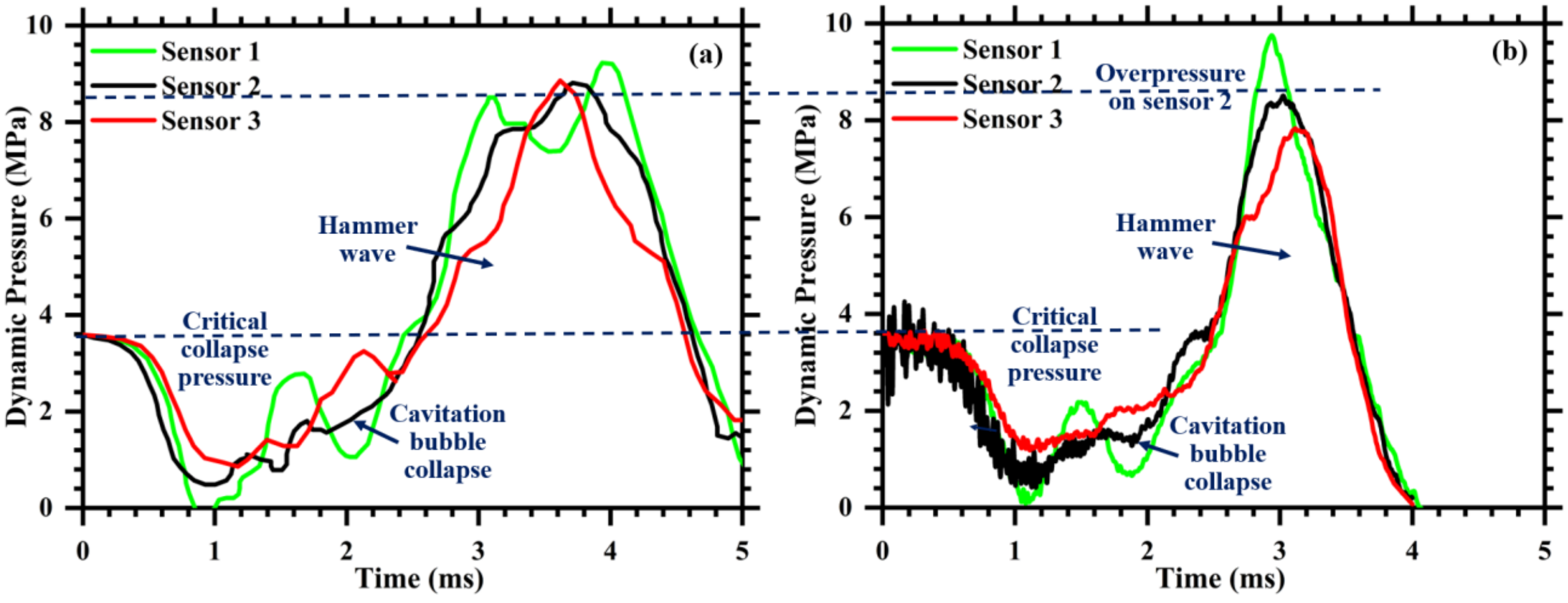

Fig. 5: Comparison of (a) experimental and (b) numerical pressure-time histories during hydrostatic implosion in a semi-confined cylinder

Immediately following collapse, both figures show the emergence of a strong water-hammer pulse. In the experiment, this occurs between 2.5–4 ms, peaking at approximately 8.52 MPa at Sensor 2. The simulation closely replicates this behavior, with the pulse appearing in the same time window and reaching a comparable peak of 8.43 MPa. This pulse exceeds critical collapse pressure by a factor of approximately 2, consistent with the Mie–Grüneisen prediction of shock-augmented pressure pulses (Zheng et al., 2023). The slight lead of the hammer wave in the simulation can be attributed to the inherent response delay of experimental sensors and falls well within expected experimental variability. Nevertheless, the alignment in both timing (within ±0.2 ms) and magnitude (within 1%) for the hammer wave indicates excellent agreement. Overall, the comparison demonstrates that the numerical model reliably captures the essential features of the implosion event, including pressure pulse formation, timing, amplitude trends, and the dynamic interaction between the fluid and structural boundaries. This close agreement validates the model's capability to simulate semi-confined implosion scenarios with high fidelity.

### 3.2 Effect of Variations in Material Properties and Geometric Parameters

Following the experimental validation of the numerical model, this section investigates how variations in material properties and geometric parameters influence the structural collapse behavior and associated fluid–structure interactions during hydrostatic implosion. The analysis focuses on the role of cylinder material, length-to-diameter ratio, and confinement geometry in governing buckling onset, energy absorption characteristics, and the nature of pressure wave propagation resulting from collapse.

#### 3.2.1 Time-resolved Kinetic Energy Absorption Comparison

Fig. 6 illustrates the time evolution of the total instantaneous kinetic energy of the entire fluid domain during hydrostatic implosion for eight semi-confined configurations, highlighting the influence of material properties and geometric parameters on energy transfer. The total instantaneous kinetic energy is computed as the spatial integral of one half times the fluid density multiplied by the square of the local velocity magnitude over all active Eulerian fluid elements at each time step. Therefore, it reflects the global kinetic energy of the surrounding water rather than a localized value at a specific position.

Although kinetic energy within the flow field is spatially non-uniform due to localized jetting, collapse asymmetry, and wave interactions, the domain-integrated kinetic energy serves as a scalar metric to quantify the overall energy transferred from the collapsing structure to the fluid. This global measure enables direct comparison of implosion severity across different materials, L/D ratios, and confinement diameters. To complement this integrated metric, spatial velocity and pressure contour plots are provided in subsequent figures to illustrate the non-uniform distribution of fluid motion and localized high-velocity regions during collapse. The labels (e.g., Al_150_2 or Ti_250_5) denote the material of the imploding cylinder (Al for aluminum, Ti for titanium), the diameter of the semi-confined enclosure in millimeters (150 or 250), and the length-to-diameter (L/D) ratio of the tube (2 or 5).

The effect of confinement diameter is evident across both materials. For a fixed L/D ratio, larger enclosures (250 mm) consistently produce sharper kinetic energy peaks. For example, Al_250_2 reaches an energy maximum of 37 J, while Al_150_2 is lower at 37 J. This pattern is similarly observed in titanium cases (Ti_250_2 vs. Ti_150_2). The higher kinetic energy peaks in larger enclosures suggest that reduced flow resistance and increased fluid mobility facilitate faster energy release during collapse, whereas smaller confinement diameters impede fluid displacement and reduce the kinetic energy release.

The influence of length-to-diameter ratio (L/D) is also significant. For a fixed confinement diameter, cylinders with L/D = 2 consistently exhibit higher peak kinetic energies and narrower curves than those with L/D = 5. For instance, Al_150_2 peaks around 30 J, while Al_150_5 peaks near 12 J. A similar trend appears for titanium: Ti_250_2 reaches approximately 118 J, whereas Ti_250_5 peaks significantly lower. This indicates that shorter, stubbier cylinders collapse more abruptly, producing more concentrated fluid acceleration, while longer tubes deform more gradually, releasing energy over a longer duration.

Comparing the two materials, titanium consistently results in higher peak kinetic energies than aluminum for equivalent geometries. For example, Ti_250_2 reaches a maximum of 118 J, nearly quadrupling the ~30 J observed for Al_250_2. This difference reflects titanium's higher elastic modulus and yield strength, which allow the structure to store more strain energy prior to collapse and release it more rapidly upon failure, leading to greater fluid acceleration.

It is important to distinguish between total fluid kinetic energy and local peak overpressure when interpreting the effect of confinement diameter. The kinetic energy curves shown in this section represent the domain-integrated fluid kinetic energy and therefore quantify the overall mass of fluid accelerated during collapse. For a fixed L/D ratio, increasing the confinement diameter increases the available fluid volume and reduces geometric blockage, allowing a larger fluid mass to participate in the motion. This can produce sharper and higher global kinetic energy peaks.

However, local water hammer amplification at a given sensor location is additionally governed by wave reflection and impedance interaction at nearby rigid boundaries. Semi-confined configurations can therefore generate amplified peak pressures relative to unconstrained conditions due to reflection-driven reinforcement, even when the total fluid kinetic energy differs. As the confinement diameter increases further toward the free-field limit, reflective amplification diminishes and local peak pressures approach unconstrained behavior. Thus, global kinetic energy amplification and local pressure amplification are related but distinct metrics governed by different confinement mechanisms.

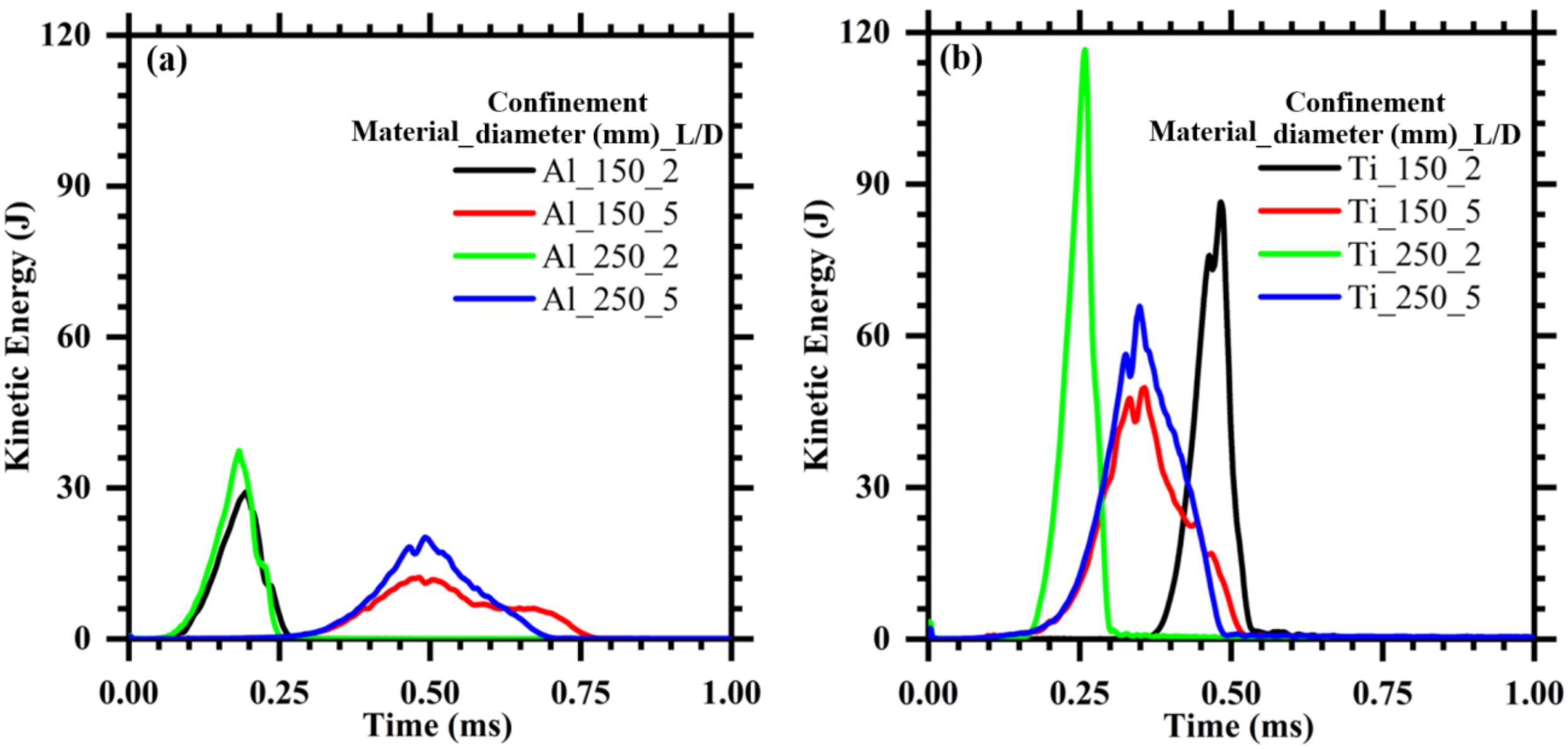

Fig. 6: Time evolution of total fluid-domain kinetic energy during hydrostatic implosion across eight semi-confined configurations. (a) Aluminum and (b) titanium cylinders are evaluated under varying confinement diameters (150 mm and 250 mm) and length-to-diameter (L/D) ratios (2 and 5). Curve labels follow the format Material_ConfinementDiameter_L/D (e.g., Al_150_2 refers to an aluminum cylinder in a semi-confined tube with a 150 mm confinement diameter and L/D = 2). The plots illustrate how material properties and geometric parameters influence the timing and magnitude of energy release, providing insight into collapse severity and fluid–structure interaction behavior in semi-confined environments.

Overall, the results demonstrate that both geometric and material parameters strongly govern the intensity and timing of energy release during implosion. Maximum kinetic energy serves as a clear indicator of implosion severity and has direct implications for subsequent fluid–structure interaction, wave propagation, and potential structural damage in surrounding components.

### 3.2.2 Time-resolved Total Stain Energy

Fig. 7 presents the evolution of total strain energy absorbed by the implodable cylinders during hydrostatic collapse, reflecting the internal work done by the structure in resisting deformation. Strain energy in this context quantifies the portion of external hydrostatic pressure converted into permanent structural deformation, primarily plastic bending, folding, and localized buckling of the cylinder walls. As collapse progresses, this energy accumulates rapidly until deformation saturates and the structure can no longer absorb further load.

Each curve label (e.g., Ti_250_2) encodes three defining parameters: the cylinder material (aluminum or titanium), the confinement diameter in millimeters (150 or 250), and the length-to-diameter (L/D) ratio (2 or 5). These parameters collectively shape how the structure stores and dissipates energy during implosion. A close inspection of the plots reveals the influence of confinement diameter on energy absorption. For both materials, wider enclosures (250 mm) generally correspond to higher final strain energy values. This is due to the fact that larger enclosures allow the fluid to move more freely, enabling the structure to deform more extensively before reaching equilibrium. The surrounding fluid in a broader confinement offers less immediate resistance, delaying collapse arrest and allowing more plastic work to accumulate.

The L/D ratio also governs how strain energy is absorbed. Cylinders with a smaller L/D tend to absorb energy quickly, often with a sharper transition to the plateau. In contrast, slender tubes (L/D = 5) take longer to reach peak energy and tend to display more gradual energy growth. This is indicative of a slower, more progressive collapse mode, where strain is distributed over a larger structural length, delaying localized energy concentration. This distinction has important implications: lower aspect ratio cylinders may produce stronger initial shock waves, while slender cylinders may exhibit greater structural stability but extended deformation periods, potentially affecting how shock is transmitted or mitigated in confined underwater systems. Moreover, the timing of peak strain energy affects design choices. Rapid peaks in low L/D cylinders indicate abrupt collapse and strong pressure transients, useful for high-impact applications but riskier for surrounding structures. Delayed peaks in high L/D cylinders reflect gradual energy release, offering better control and shock mitigation. Therefore, low L/D is preferred for energy transfer, while high L/D is better for structural stability and reduced shock impact in sensitive environments. When comparing materials, the contrast is striking. Titanium cylinders absorb significantly more energy than aluminum ones under all comparable conditions. This stems from titanium's higher elastic modulus and yield strength, which allow it to resist deformation more effectively. As a result, it stores more energy internally before structural failure occurs. For example, while aluminum cylinders level off between 330–390 J, titanium cylinders exceed 900 J in some configurations, more than double in many cases. The highest value is observed in Ti_250_5, which reaches approximately 960 J. In contrast, Al_250_5 peaks near 390 J. This difference underscores titanium's ability to undergo more substantial plastic work under the same external pressure conditions. In terms of design choice, this makes titanium more suitable for applications requiring high energy absorption and structural robustness, especially in high-pressure or impact-critical environments. In contrast, aluminum may be preferred where lower energy demands, weight savings, or cost efficiency are prioritized.

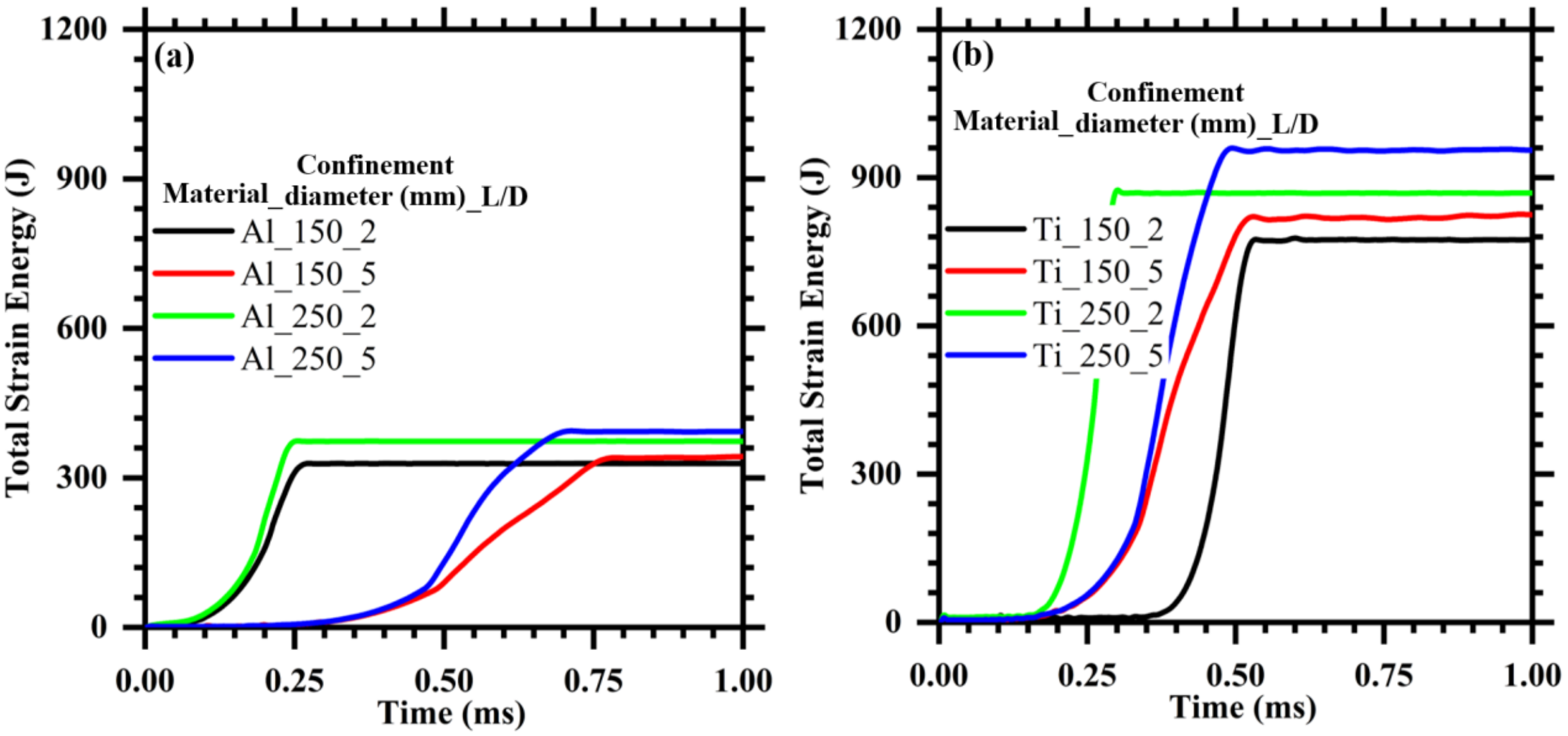

Fig. 7: Time evolution of total strain energy absorption during hydrostatic implosion for (a) aluminum and (b) titanium cylinders under semi-confined conditions. Each curve is labeled as Material_Confinement Diameter_L/D, where "Material" indicates aluminum (Al) or titanium (Ti),

“Confinement Diameter” is the inner diameter of the surrounding enclosure in millimeters (150 mm or 250 mm), and “L/D” is the length-to-diameter ratio of the implodable cylinder (2 or 5). The plots illustrate how material properties and geometric parameters influence the magnitude and rate of strain energy accumulation during collapse.

Although material properties, such as yield strength and elastic modulus, strongly govern the total amount of strain energy absorbed during collapse, the timing of when this energy is absorbed is not dictated by material alone. Instead, it results from a coupled interaction between material behavior and geometric slenderness, characterized by the length-to-diameter (L/D) ratio of the cylinder. In low L/D configurations, the structural response is more dominated by geometric stiffness. The shorter axial length offers limited space for progressive deformation, causing the collapse to localize rapidly once the critical pressure is reached. For softer materials like aluminum, this means that deformation initiates early and proceeds quickly, resulting in earlier peaks in strain energy despite their lower energy absorption capacity. The collapse is more impulsive, but the structure cannot resist long enough to accumulate significant internal work. In contrast, high L/D configurations (slender cylinders) introduce more complex deformation patterns. Here, buckling initiates more gradually and propagates along the length of the tube. For stiffer materials like titanium, this extended geometry allows the structure to engage more of its length in deformation, and the higher yield strength resists collapse longer before instability grows. Once the collapse initiates, however, the rate of energy accumulation accelerates sharply, especially in titanium, due to the rapid release of stored elastic energy over a larger deforming region. This explains why titanium cylinders in high L/D configurations reach peak energy faster than aluminum, despite their inherent strength, their collapse is delayed but more energetic and abrupt once triggered. Therefore, the timing of the peak is not a material property alone, it reflects how material strength, stiffness, and collapse mechanics interact with geometry. The L/D ratio modulates the deformation mode (localized vs. distributed), the collapse duration, and ultimately, the strain energy growth rate. This coupling is critical for predicting implosion behavior and for designing structures where the timing of energy release (e.g., for shock management or containment) is just as important as total energy absorption. In design terms, the coupled effect of material and L/D ratio means choices must be application-specific. Titanium with high L/D is ideal for maximizing energy absorption and structural resilience, but it leads to more abrupt, high-intensity collapse. Aluminum with low L/D collapses earlier and more gently, making it better for minimizing shock transmission.

### 3.2.3. Dynamic pressure-time responses

This section investigates dynamic pressure-time responses during hydrostatic implosion for various material and geometric configurations under semi-confined conditions. Each configuration is analyzed individually to reveal how confinement diameter, material type, and slenderness ratio (L/D) affect pressure waveforms and fluid–structure interaction. The dynamic pressure time histories are accompanied by inset images showing pressure contours in the imploding cylinders at key moments, providing spatial insight into localized stress concentrations and collapse progression. In this subsection, the plotted data in the figures are extracted from the sensor located at position 2 within the simulation setup.

It is important to note that the pressure histories presented here correspond to the local hydrodynamic pressure obtained directly from the LS-DYNA S-ALE solution at the specified monitoring locations. In the adopted formulation, pressure is computed through the equation of state and represents the thermodynamic pressure of the fluid, including compressibility and shock contributions during the implosion process. These values are recorded directly from the solver output at each time step and do not involve any post-processing based on velocity fields. It should therefore be emphasized that the plotted pressure should not be interpreted in the classical fluid mechanics sense based on Bernoulli's equation. Instead, the results represent transient hydrodynamic pressure responses associated with collapse-induced wave propagation, cavitation, and water hammer effects.

Fig. 8 presents an initial rapid rise in dynamic pressure immediately after collapse initiation, with both cases collapsing into mode 4. In Fig. 8a, the aluminum cylinder (L/D = 2) confined within a 150 mm enclosure shows the first pressure spike escalates to approximately 31 MPa within about 0.3 ms, indicating a strong water-hammer event. This spike is followed by cyclical pressure fluctuations between 20 and 30 MPa over the next 5 ms, reflecting the confinement-induced reflections. In comparison, Fig. 8b shows the same aluminum cylinder (L/D = 2) within a wider 250 mm semi-confined tube, where the first overpressure peak reaches around 33 MPa at approximately 0.2 ms, slightly higher and earlier than in the smaller confinement. The subsequent oscillations are of lower amplitude, stabilizing around 25–27 MPa by the end of the 6-ms window.

The water hammer effects illustrated in the dynamic pressure–time histories for the two configurations in Fig. 8 exhibit notable differences in amplitude, persistence, and energy dissipation, all of which are critical indicators of FSI dynamics during hydrostatic implosion. Beyond the initial peak, the differences become more pronounced. In Fig. 8a, the dynamic pressure response features a series of sustained oscillations that persist throughout the 6ms. These secondary peaks and troughs indicate continued wave reflections and interactions between the collapsing structure and the fluid domain. This sustained oscillatory behavior reflects strong fluid–structure coupling and a slower rate of energy dissipation. This is corroborated by the insets showing localized high-pressure zones on the deforming cylinder, suggesting that complex buckling patterns may continuously excite secondary pressure waves.

By contrast, Fig. 8b shows a distinctly different trend. While the initial water hammer is higher, the subsequent oscillations are heavily damped. By around 1.5 ms, the dynamic pressure stabilizes and fluctuates within a narrow range, with an average value hovering around 25 MPa. The faster decay of oscillations implies that energy from the initial shock was more rapidly dissipated, potentially through smoother plastic deformation or more homogeneous structural collapse. The pressure field in the insets for 8 (b) supports this, revealing more uniformly distributed pressure with fewer localized spikes compared to 8 (a). The inserted contour plots highlight localized pressure fields on the collapsing cylinder surfaces during peak events: the near-wall regions exhibit pressures ranging from approximately -155 MPa, which is the cavitation zone, to +190 MPa, which is the impact zone, depending on geometry. Fig. 8b shows higher localized pressure extremes compared to Fig. 8a, correlating with the larger observed peak in the pressure history.

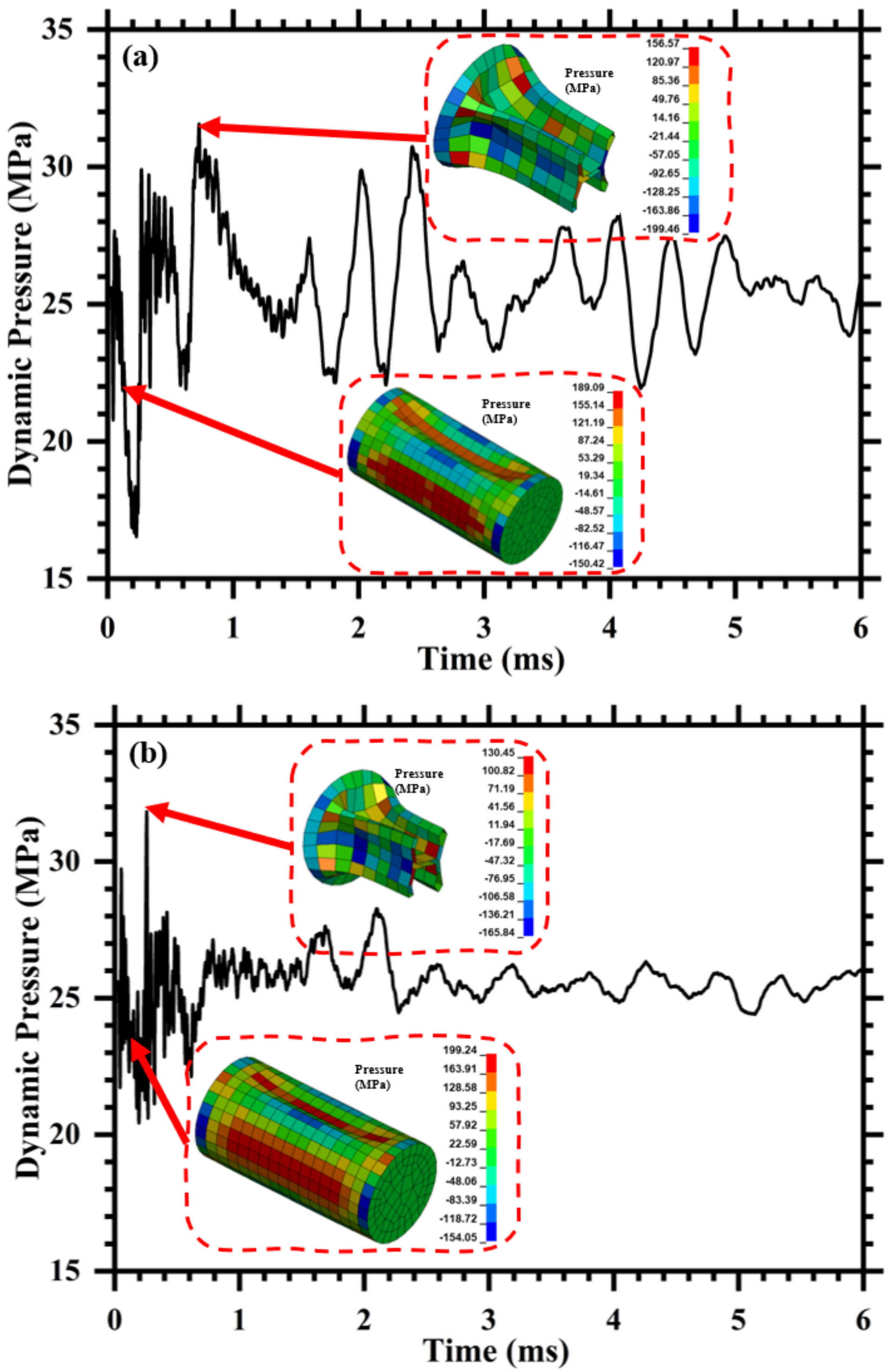

Fig. 8: Dynamic pressure response during hydrostatic implosion of aluminum cylinders with L/D ratio of 2 under semi-confined conditions: (a) 150 mm confinement diameter, (b) 250 mm confinement diameter. Each plot shows pressure histories recorded at sensor position 2, with inset contour images depicting pressure distributions on the cylinder surface at peak collapse. The results illustrate the influence of confinement diameter on the magnitude, timing, and damping behavior of water-hammer pressure spikes.

While Fig. 8 focused on aluminum cylinders with L/D = 2, Fig. 9 presents results for the same material but with a L/D = 5, allowing for a direct comparison of how slenderness ratio influences dynamic pressure evolution during hydrostatic collapse. Compared to the L/D = 2 configurations in Fig. 8, which exhibited mode 4 buckling and sharper pressure spikes, these L/D = 5 cases in Fig. 9 show mode 2 buckling with significantly delayed collapse initiation, lower peak overpressures, and more subdued oscillatory behavior. This difference in buckling mode plays a central role: higher-order modes (like mode 4) induce more abrupt, localized deformation, triggering stronger fluid impulses, while lower-order modes (like mode 2) are more stable and spread out, resulting in reduced and delayed water hammer effects. These observations underscore how both slenderness ratio and buckling mode shape influence the nature of fluid–structure interactions and the intensity of transmitted shock in semi-confined underwater implosion scenarios.

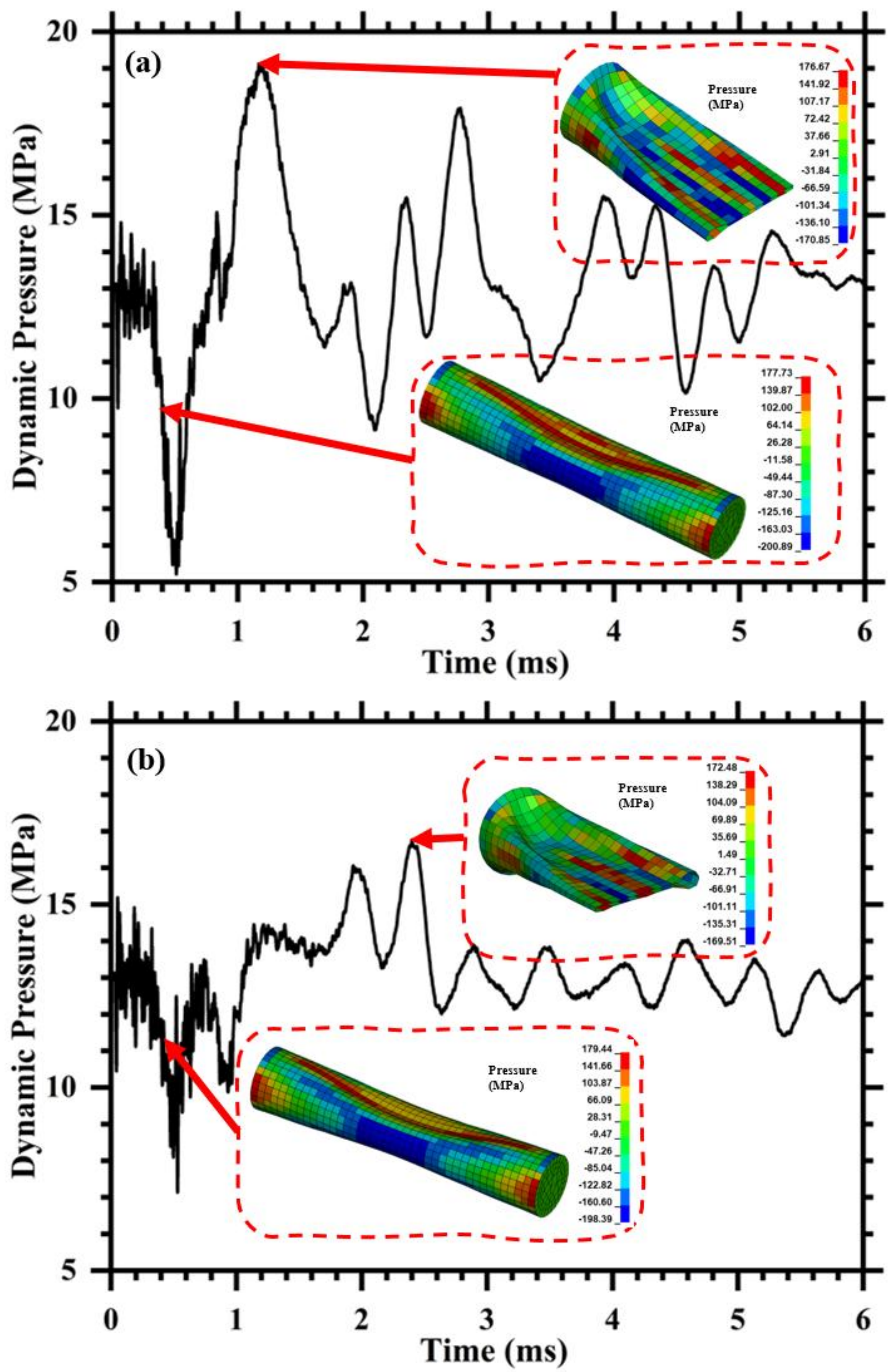

Fig. 9: Dynamic pressure response during hydrostatic implosion of aluminum cylinders with L/D ratio = 5 under semi-confined conditions: (a) 150 mm confinement diameter, (b) 250 mm confinement diameter. Pressure histories are obtained from sensor position 2. Inset contour plots

show internal pressure distributions at key collapse stages, highlighting differences in collapse severity and fluid–structure interaction due to variation in confinement diameter.

In Fig. 9a, pressure abruptly transitions from the hydrostatic baseline of critical collapse pressure of 13.1 MPa to a trough near 5 MPa at around 0.25 ms, signifying rapid cavitation as the wall implodes inward. This is followed by a prominent first overpressure pulse, peaking at approximately 19 MPa, a 1.5× increase over the hydrostatic load, occurring at around 1.1 ms. Subsequent overpressure oscillations persist for nearly 5 ms, decaying gradually to around 13 MPa, indicative of repeated impedance reflections within the confined volume. Fig. 9b features a pressure dip at approximately 7 MPa at 0.5 ms, followed by a similar first overpressure peak reaching 17 MPa at around 1.2 ms. The broader temporal spacing between cavitation trough and overpressure peak compared to Fig. 9a suggests slower fluid motion and collapse progression due to the larger confinement diameter. Post-primary pulse, pressure oscillations again decay toward 13 MPa over the succeeding milliseconds, but with a slightly smoother waveform.

Embedded contour insets of the mode 2 deformed cylinders in Fig. 9 show localized pressures acting on the cylinder surface during peak events, with tensile pressures rising to approximately 175 MPa and compressive stresses falling near −170 MPa, highlighting intense elastic–plastic stress levels preceding buckling. These stress extremes emphasize the role of geometric confinement in amplifying structural stress responses. The first overpressure pulse in both configurations manifests at nearly the same amplitude, but with a 1.1 ms delay for the larger confinement diameter, confirming that confinement geometry influences collapse timing more strongly than magnitude. The water hammer responses observed in the two configurations in Fig. 9 reveal contrasting FSI characteristics during hydrostatic collapse. In Fig. 9 (a), the initial water hammer manifests as a pronounced pressure spike. This peak is followed by a series of damped oscillations that gradually decay but maintain significant amplitude up to 5 ms. The persistence of these oscillations suggests a strong coupling between the collapsing structure and the surrounding fluid, with pressure waves continuously reflecting within the semi-confined domain. In contrast, Fig. 9 (b) exhibits a water hammer of lower intensity. Although the pressure-time history also displays oscillatory behavior, the amplitude of these oscillations is relatively subdued and decays more rapidly compared to Fig. 9 (a). The first sharp pressure drop occurs earlier and is less severe, indicating a more progressive and possibly more axisymmetric deformation mode.

While Figs. 8 and 9 presented the dynamic pressure responses of aluminum cylinders, Fig. 10 shifts focus to titanium cylinders under identical L/D = 2 conditions. The contrast in material behavior is immediately evident in the pressure-time histories and associated deformation modes, with titanium showing sharper and more intense fluid–structure interaction signatures due to its higher stiffness and strength. In Fig. 10a, the hydrostatic critical collapse pressure of approximately 56 MPa rapidly transitions into a downward spike reaching 43 MPa around 0.4 ms, indicative of localized cavitation as the fluid rushes inward. A sharp first overpressure arrives at 0.8 ms, peaking at 63 MPa, a surge roughly 1.2 times above the critical collapse pressure. This is immediately followed by pressure oscillations that stabilize around 55 MPa, reflecting reflections within the confined vessel. In contrast, Fig. 10b yields an even higher initial overpressure peak of approximately 74 MPa around 0.7 ms, roughly 1.23 times the critical collapse pressure before settling into a sustained pressure plateau near 60 MPa through 6 ms.

The inset pressure contour maps further illustrate the stress dynamics. In Fig. 10b, the cylinder structural surface experiences extremely high compressive pressures exceeding –400 MPa and tensile pressures above +350 MPa localized near buckling lobes. These stress extremes corroborate the magnitude of the recorded pressure pulses, confirming that the L/D ratio of 2 and titanium material for the cylinder amplify both fluid pressure peaks and structural stress.

The water hammer overpressure in Fig. 10a initiates with a sharp and intense pressure surge. This is immediately followed by a steep pressure drop and multiple oscillatory spikes, with amplitudes fluctuating over a 5 ms duration. The persistence and periodicity of these oscillations suggest a highly energetic interaction between the fluid and the collapsing structure. The pressure contours illustrate a pronounced inward buckling near the mid-span and significant localized plastic deformation at the end caps, confirming a combination of axisymmetric collapse and local folding. In contrast, Fig. 10b presents a more abrupt collapse and smoother decay, indicating a faster energy transfer and damping within the system. The deformation mode varies between the two configurations. In Fig. 10a, the cylinder undergoes mode 3 buckling, with three lobes forming around the circumference, whereas Fig. 10b displays mode 4, with four lobes. The shift to a higher buckling mode in the 250 mm confinement arises from the increased fluid interaction area and titanium's mechanical properties namely, its high strength-to-weight ratio and reduced plastic capacity, which favor more symmetric and finely distributed buckling patterns.

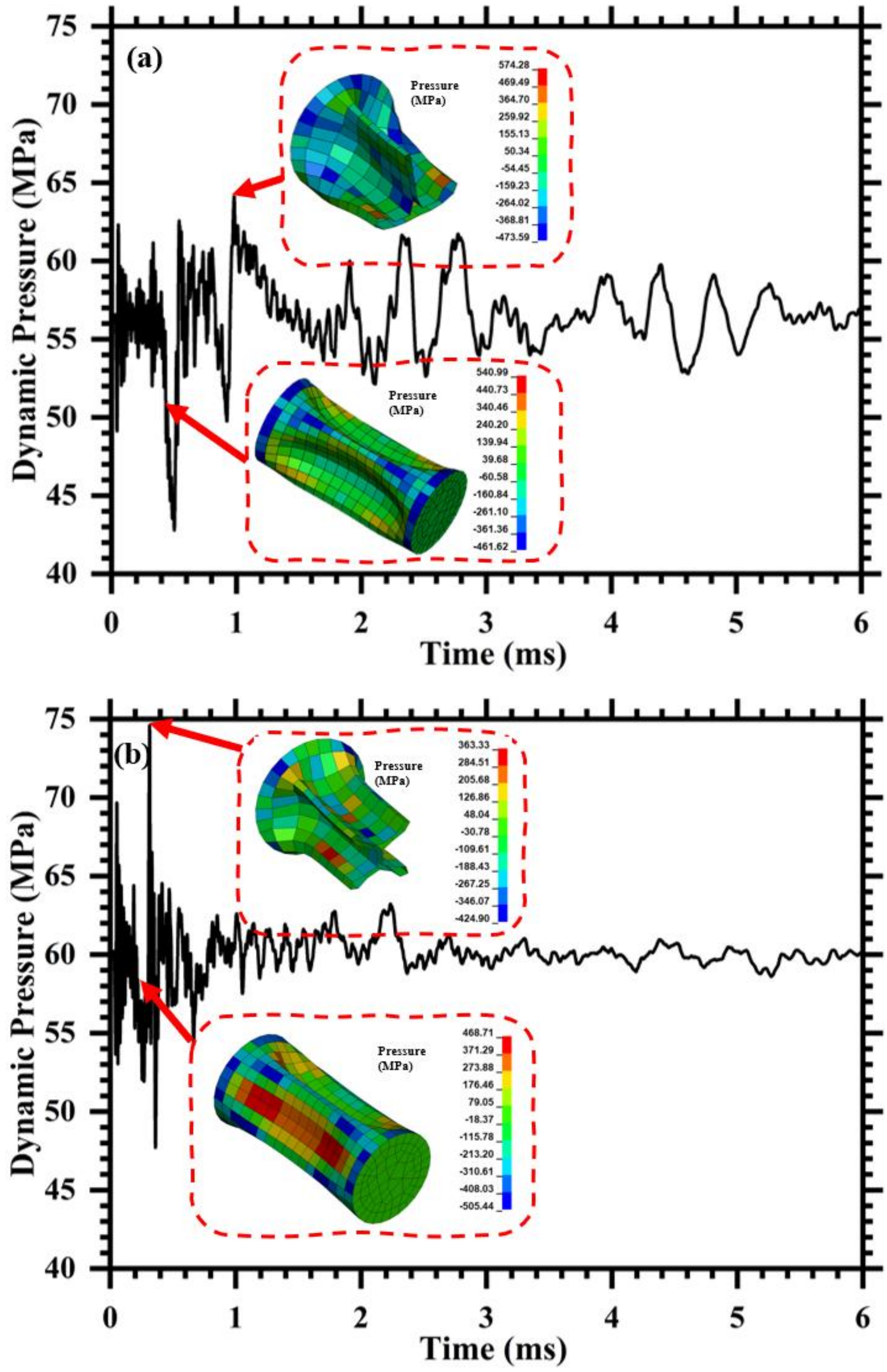

Fig. 10: Dynamic pressure time histories for hydrostatic implosion in semi-confined titanium configurations with L/D = 2. (a) Semi-confinement diameter = 150 mm, (b) 250 mm. Insets show internal pressure contours on the imploding cylinder at key collapse stages.

Fig. 11 presents dynamic pressure time histories for hydrostatic implosion in semi-confined titanium cylinders with a slenderness ratio of L/D = 5. In both Fig. 11(a) and (b), the collapse initiates from a hydrostatic critical pressure of 28.82 MPa, which is quickly followed by a sharp pressure drop to approximately 15 MPa within 0.5 ms. This rapid decline is attributed to fluid acceleration and cavitation triggered by the inward structural collapse. In Fig. 11a, the first overpressure peak reaches about 38 MPa at roughly 0.9 ms, representing a nearly twofold increase over the hydrostatic critical collapse pressure. Dynamic oscillations then decay gradually toward 30 MPa over the next 5 ms. The embedded contour maps of a mode 2 deformation reveal that surface pressures during collapse span from +570 MPa in tensile to –250 MPa in compressive, indicating extreme local stress concentrations.

Fig. 11b, representing a larger semi-confinement diameter, shows a delayed first pulse at approximately 2.2 ms, with a slightly reduced peak pressure of ~33 MPa. The subsequent oscillations exhibit lower amplitude and reduced variability, stabilizing again near 30 MPa. The inset pressure maps show a narrower stress range, –250 MPa to +430 MPa, suggesting that larger confinement moderates the collapse intensity while slightly prolonging its timeline.

In both Fig. 11(a) and 11(b), the collapse event initiates a rapid pressure surge, characteristic of a primary water hammer, as the structural failure instantaneously displaces the surrounding fluid and generates strong compressive waves. In Fig. 11 (a), the primary water hammer peak is followed by a series of well-defined, high-amplitude oscillatory waveforms that persist up to 5 ms, suggesting sustained pressure wave reflections and interactions within the semi-confined fluid domain. The intensity and frequency of the oscillations reflect a structure-fluid system with energy transmission pathways, possibly due to a combination of geometry and material compliance. Conversely, Fig. 11(b) shows a more subdued water hammer phenomenon consistent with results from Figs 8, 9, and 10. The initial pressure spike is followed by pressure oscillations of noticeably reduced amplitude. After the initial collapse, the system transitions into a lower-intensity response regime, with dynamic pressure. The dampened oscillatory pattern implies either a more dissipative interaction within the fluid domain or a structure with higher damping characteristics that reduces the effectiveness of wave reflections and resonance buildup.

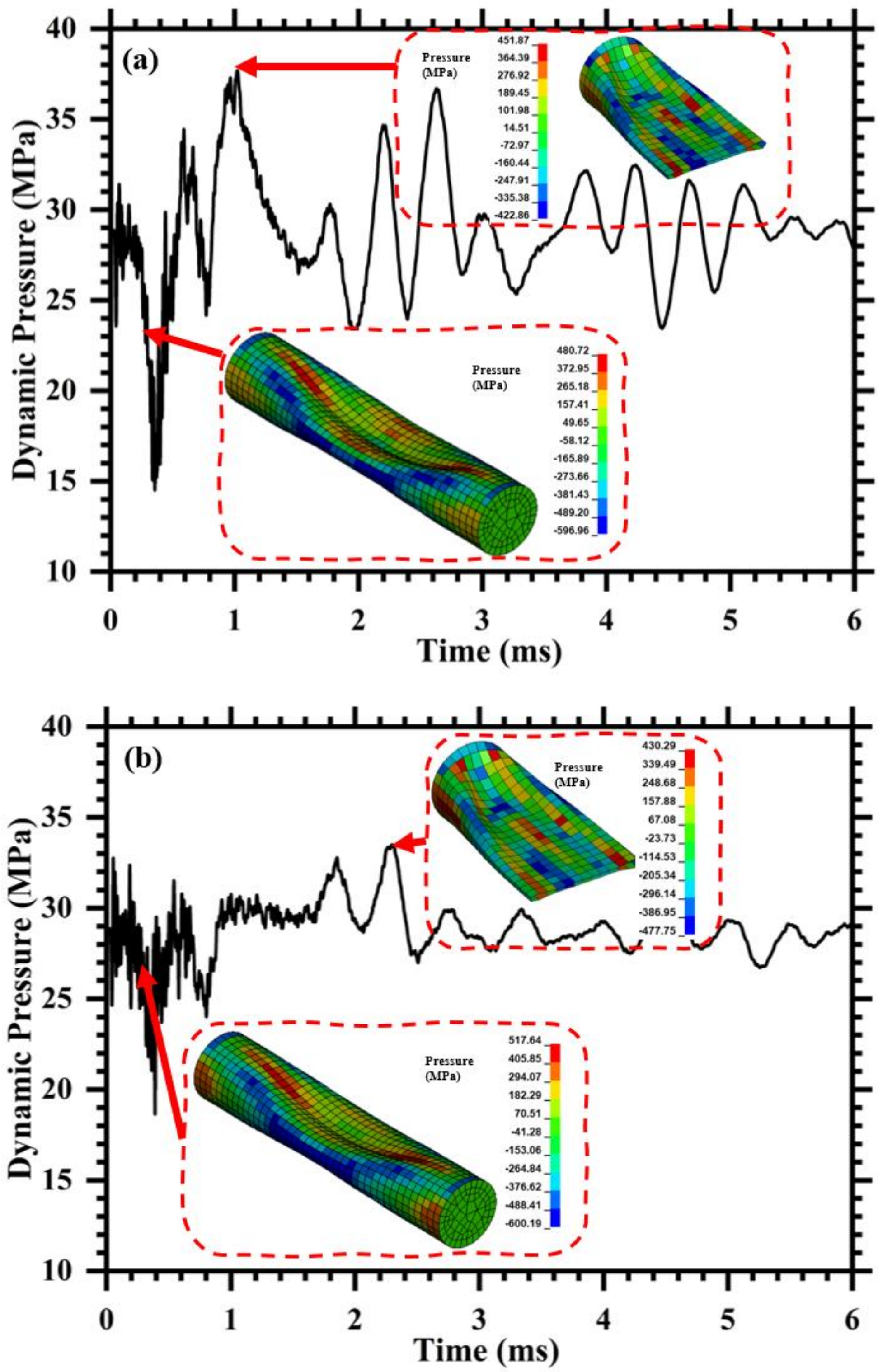

Fig. 11: Dynamic pressure response during hydrostatic implosion in semi-confined configurations using titanium cylinders with L/D = 5. (a) Confinement diameter = 150 mm, (b) Confinement diameter = 250 mm. Insets illustrate internal pressure contours on the cylinder surface at key collapse stages, highlighting pressure distributions and buckling deformation.

Overall, Figs. 8–11 reveal clear trends in how material type, L/D ratio, and confinement diameter affect hydrostatic implosion behavior. Titanium cylinders consistently exhibit higher peak pressures, more intense water hammer effects, and stronger pressure oscillations than aluminum, due to their greater strength and stiffness. Lower L/D ratios (L/D = 2) lead to higher-order buckling modes and earlier, sharper pressure spikes, while higher L/D ratios (L/D = 5) result in mode 2 deformation with more gradual collapse and damped pressure profiles. Increasing the confinement diameter generally delays collapse onset and shifts pressure peaks later, with varying effects on amplitude depending on the material. Overall, material and geometry interact in a coupled way to shape the fluid–structure response, highlighting the need to balance strength, mass, and energy dissipation based on the intended application, whether for maximizing shock resistance or minimizing transmitted loads in confined underwater systems.

### 3.2.4. Fluid-structure interaction average pressure-time responses

While the previous section analyzed dynamic pressure-time responses from discrete sensors, this section leverages fluid–structure interaction (FSI) average pressure-time plots to provide a more spatially and temporally resolved picture of the implosion process. These plots reveal critical aspects of the collapse not captured by scalar pressure data, such as the evolution of fluid velocity fields, asymmetric jetting, localized cavitation, and vortex structures. They also visualize how collapse fronts propagate along the cylinder, how structural buckling modes evolve, and how fluid motion feeds back into structural deformation. The ability to capture these interactions over the full wetted surface enables a richer interpretation of stress wave reflections, energy redistribution, and transient loading patterns, thus offering high-fidelity insight into the mechanisms governing failure and fluid momentum exchange during implosion.

Fig. 12 presents a comparative analysis of the average dynamic pressure-time response and resultant fluid velocity fields during the hydrostatic implosion of aluminum cylinders in semi-confined environments at a fixed L/D = 2. The two cases correspond to different confinement diameters, specifically 150 (Fig. 12a ) mm and 250 mm (Fig. 12b), which provide insight into the influence of geometric confinement on FSI behavior. In the 150 mm diameter case shown in Fig. 12a, the average dynamic pressure rapidly decreases from an initial value of 26 MPa to a minimum of roughly 3.8 MPa at 0.23 ms. This sharp pressure drop indicates the onset of structural collapse and a subsequent rapid release of stored elastic and plastic energy into the surrounding fluid domain. The embedded velocity contour indicates a peak resultant fluid velocity of approximately 300.7 m/s. The velocity field reveals symmetric outward jets and lateral ejection zones centered along the cylinder’s midspan, highlighting strong radial momentum transfer due to symmetric collapse. In Fig. 12b, corresponding to the 250 mm confinement diameter, the dynamic pressure exhibits a similar initial decay from 26 MPa to a minimum of 1.6 MPa at around 0.24 ms. However, the recovery in pressure is steeper, and the oscillation amplitude in the post-collapse phase is slightly higher compared to the 150 mm case. The increased fluid volume results in a more distributed energy coupling, which is reflected in the higher resultant fluid velocity reaching approximately 365.9 m/s. The velocity field shows more pronounced localized jetting, especially at the cavity neck and base regions of the collapsing structure, indicating stronger inertial interactions due to increased confinement space. The comparison illustrates that increasing the confinement diameter enhances the average peak fluid velocity and sustains longer and more dynamic pressure oscillations due to larger fluid mass engagement.

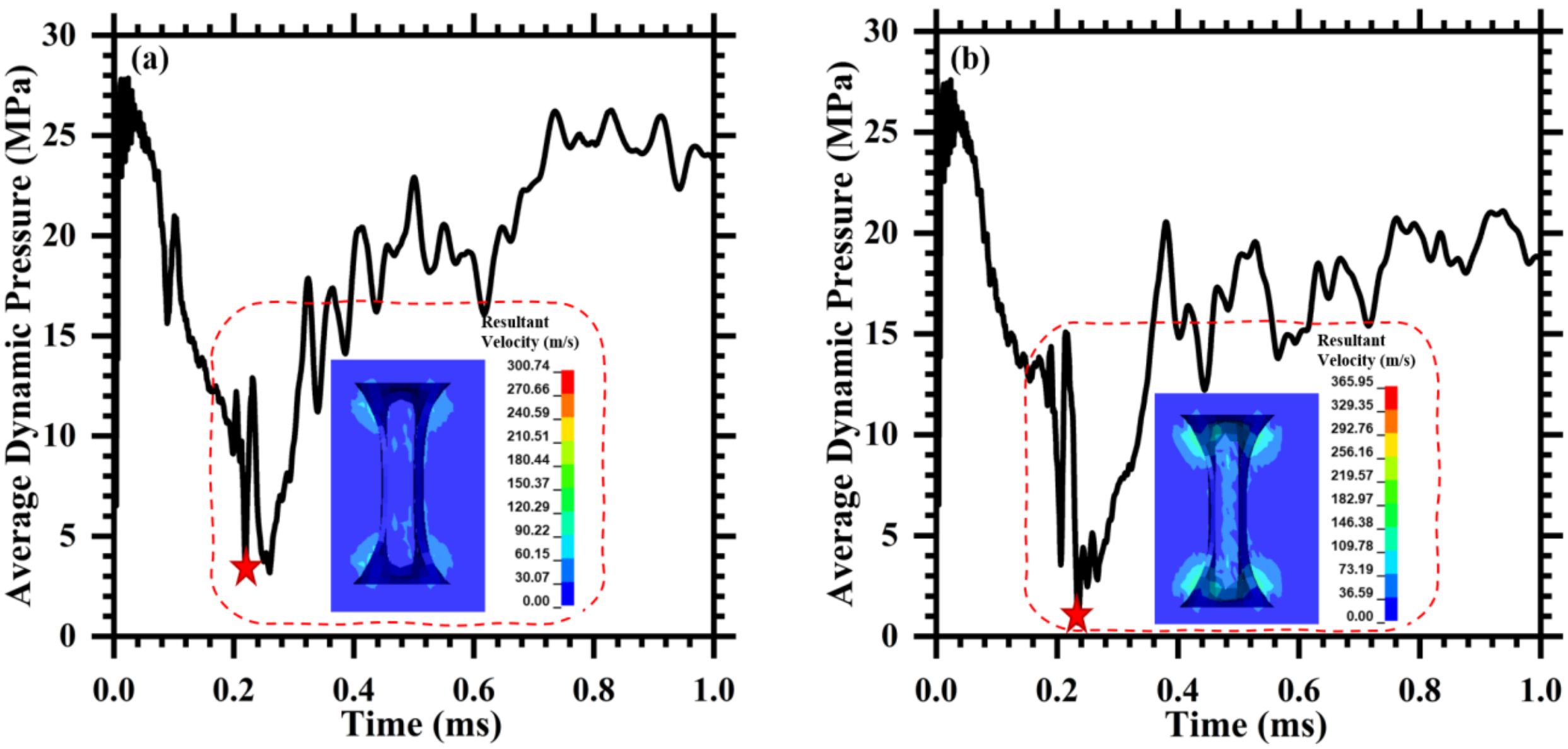

Fig. 12: Fluid-structure interaction average pressure-time response and fluid velocity field evolution for an aluminum cylinder at L/D = 2 under hydrostatic collapse for confinement diameters of (a)150 mm, (b) 250 mm.

Fig. 13 illustrates the average dynamic pressure response and fluid velocity field evolution during the hydrostatic implosion of aluminum cylinders with L/D = 5 in semi-confined configurations. Fig. 13 (a) represents the 150 mm diameter semi-confinement, while Fig. 13 (b) corresponds to the 250 mm diameter case. The collapse dynamics and resulting pressure oscillations provide critical insight into the strength and evolution of fluid-structure interaction coupling under differing confinement geometries. In Fig. 13a, the initial average dynamic pressure begins at approximately 13 MPa and experiences a rapid decay to a minimum of 4.9 MPa at 0.47 ms, marked by the red star. This substantial pressure drop signifies the moment of peak structural deformation and momentum transfer into the fluid. The embedded velocity contour reveals a maximum resultant fluid velocity of 275.9 m/s, concentrated around the midspan of the imploding cylinder. The symmetric nature of the flow field indicates a relatively uniform collapse pattern with radial expansion of the fluid driven by the cylinder's inward buckling and plastic deformation.

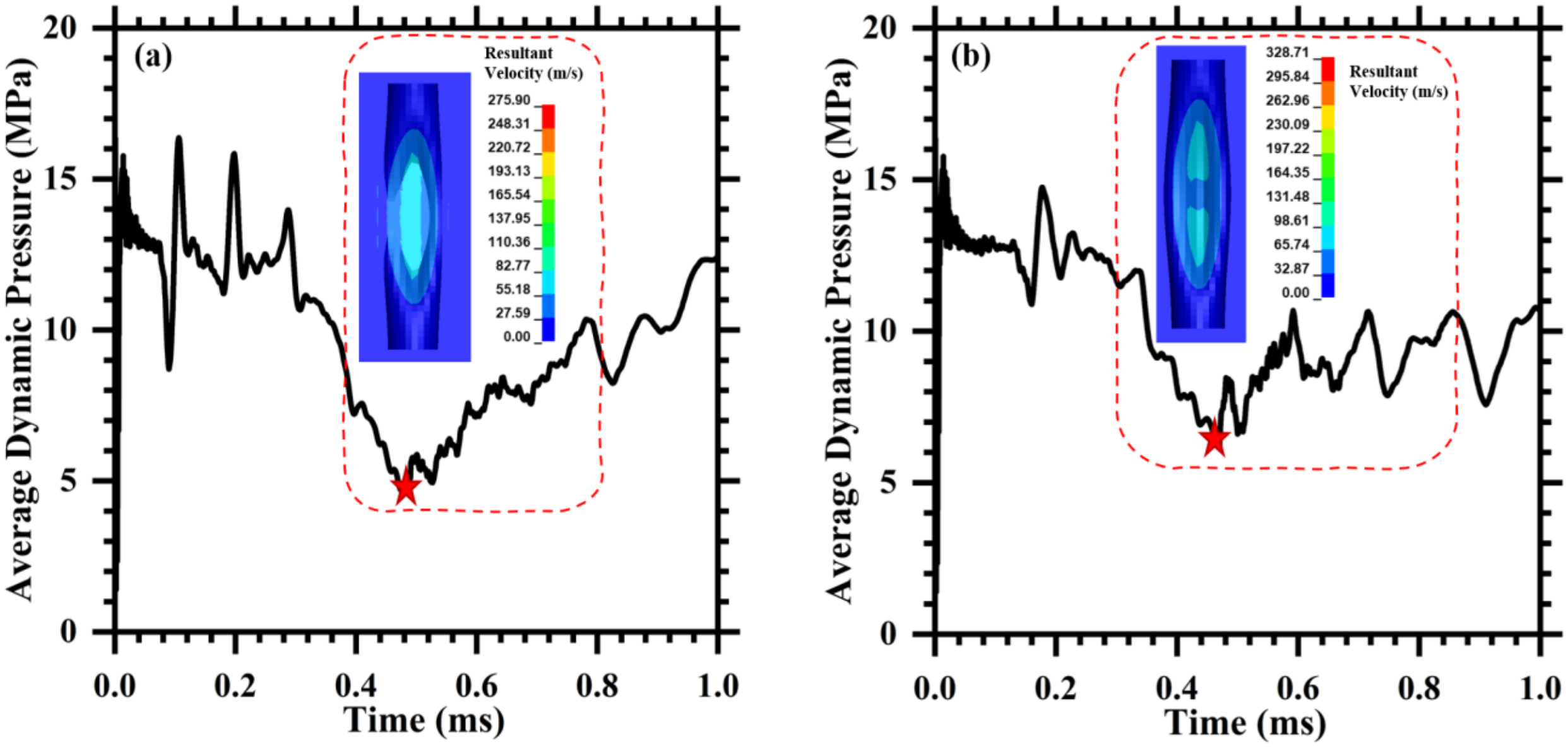

Fig. 13: Fluid-structure interaction average pressure-time response and fluid velocity field evolution for an aluminum cylinder at L/D = 5 under hydrostatic collapse for confinement diameter of (a)150 mm, (b) 250 mm.

In contrast, Fig. 13b shows a slightly more attenuated pressure response for the 250 mm confinement diameter case. The average dynamic pressure decreases from an initial value of 13 MPa to a minimum of 5.4 MPa at 0.46 ms. Although the pressure dip is slightly less abrupt than in the 150 mm case, the peak resultant fluid velocity is noticeably higher, reaching 328.7 m/s. This increase in fluid velocity is attributed to the larger volume of fluid within the confinement, allowing for greater momentum propagation and energy transfer during collapse. The velocity field highlights more localized zones of acceleration and jetting, particularly in the axial direction, suggesting a more concentrated implosion region and likely more severe cavitation near the midbody. Observations confirm that increasing the diameter of the confinement chamber leads to enhanced velocity magnitudes and prolonged oscillations, indicative of stronger fluid-structure coupling. The sharpness of the pressure troughs and recovery curves, in combination with spatial velocity field data, emphasizes how geometric parameters such as confinement diameter and L/D ratio directly govern collapse dynamics and fluid energy redistribution.

Fig. 14 presents the average dynamic pressure response and fluid velocity field evolution for titanium cylinders under hydrostatic collapse at L/D = 2, with varying semi-confinement diameters of 150 mm and 250 mm. The results highlight the pronounced influence of cylinder geometry on the strength of fluid-structure interaction, collapse intensity, and resultant fluid motion. In Fig. 14a, corresponding to the 150 mm confinement diameter case, the initial average dynamic pressure reaches a peak of 58 MPa. A sharp decay occurs shortly after, dropping to 23 MPa at around 0.48 ms, marked by a red star. This point signifies a critical moment of structural collapse and energy transfer to the fluid. The accompanying velocity field shows a maximum resultant velocity of 510.87 m/s. The velocity distribution indicates a highly asymmetric collapse profile, with fluid acceleration concentrated along the lateral flanks of the cylinder. This suggests a bending-induced failure mode that directs momentum radially outward, intensifying fluid flow within confined zones. In contrast, Fig. 14b shows the behavior for a 250 mm diameter semi-confinement. The initial pressure similarly peaks at 58 MPa but decays more rapidly to a local minimum of approximately 9 MPa at 0.36 ms. However, the resultant velocity field reaches a much higher peak value of 720.51 m/s. This dramatic increase in fluid speed is attributed to the larger internal volume of the confining fluid region, allowing for more vigorous fluid acceleration and jetting as the titanium structure collapses. The flow pattern is more symmetric and concentrated toward the axial midsection, indicating a dominant buckling mode leading to internal void formation and subsequent pressure redistribution.

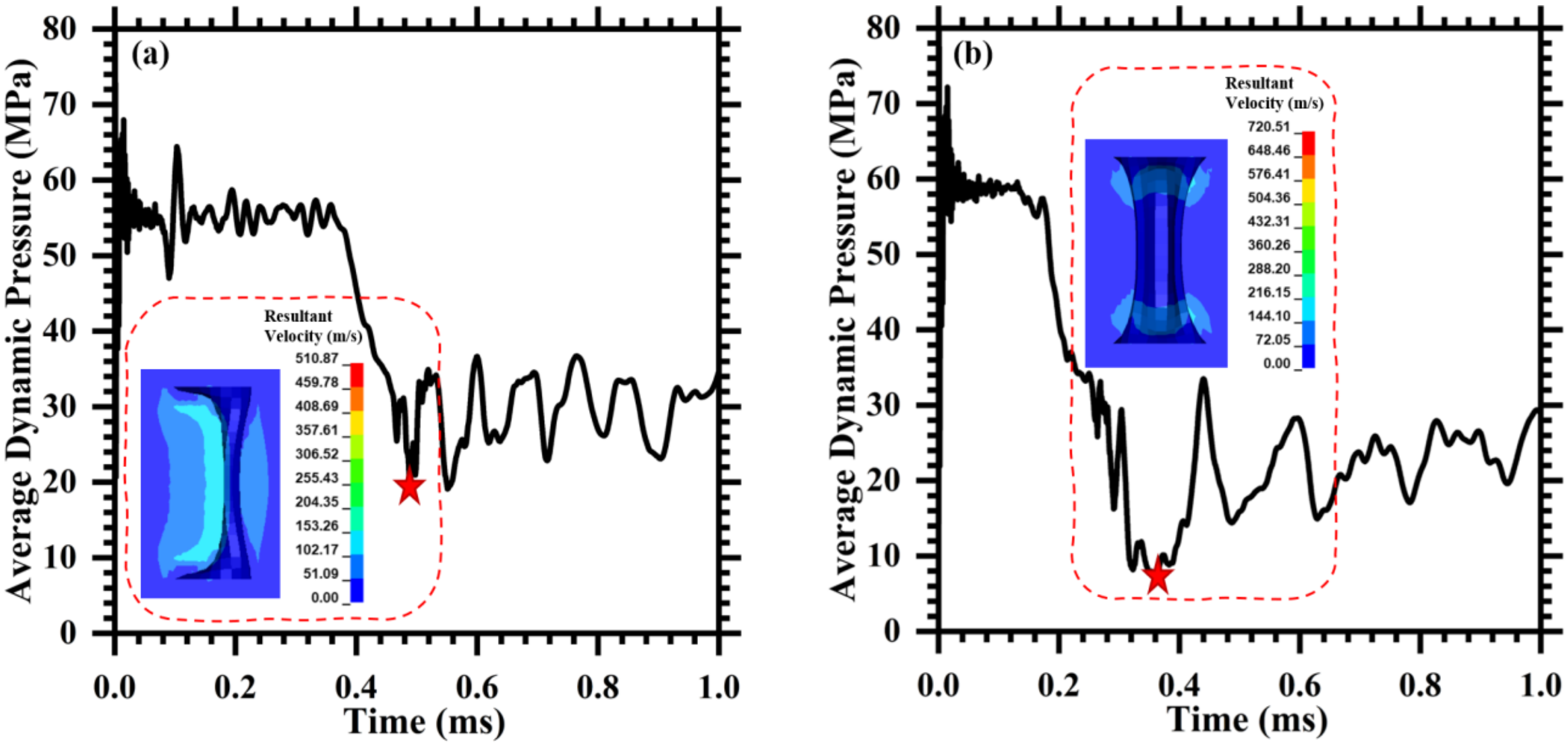

Fig. 14: Fluid-structure interaction average pressure-time response and fluid velocity field evolution for a titanium cylinder at L/D = 2 under hydrostatic collapse for confinement diameters of (a)150 mm, (b) 250 mm.

Fig. 15 presents an analysis of average dynamic pressure-time responses and corresponding fluid velocity fields during hydrostatic collapse of titanium cylinders at L/D = 5 for the two semi-confined configurations: one with a 150 mm diameter confinement (a) and the other with a 250 mm diameter confinement (b). In Fig. 15a, which corresponds to the 150 mm diameter semi-confinement, the pressure profile initiates at approximately 31 MPa and experiences multiple oscillatory fluctuations before reaching a distinct minimum near 13 MPa at 0.38 ms. This time point, marked with a red star, signifies the moment of peak structural collapse and maximum energy transfer to the fluid. The corresponding fluid velocity field reveals a maximum resultant velocity of 373.28 m/s, with the highest velocity concentrated symmetrically around the cylinder's midsection. The velocity gradient suggests that the internal collapse propagates along both the radial and axial directions, inducing symmetric vortical flows that reinforce inward pressure decay. In Fig. 15b, representing the 250 mm semi-confinement case, the initial pressure is comparable, peaking near 31 MPa. However, the pressure drops more rapidly and reaches a deeper local minimum of approximately 14 MPa at 0.38 ms. Despite the similar minimum pressure, the larger confinement volume enables a more pronounced fluid response, evidenced by the resultant velocity peak of 406.51 m/s. The velocity field visualization shows more defined elliptical isovelocity zones, indicating the formation of localized high-velocity jets during the structural inward buckling. These fluid jets are more diffused than in the smaller confinement case, suggesting an increased capacity for energy redistribution within the fluid domain.

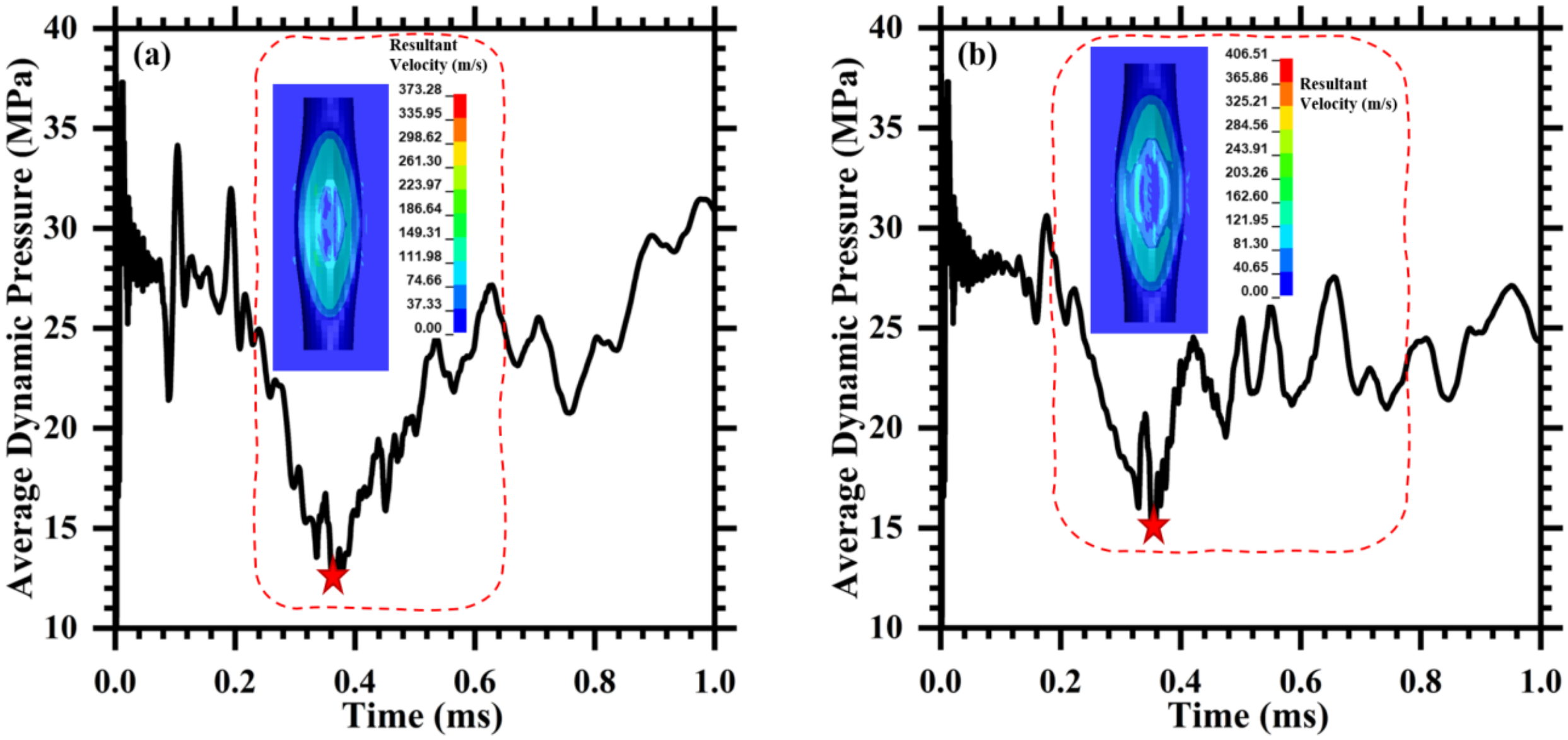

Fig. 15: Fluid-structure interaction average pressure-time response and fluid velocity field evolution for a titanium cylinder at L/D = 5 under hydrostatic collapse for (a)150 mm diameter, (b) 250 mm diameter of semi-confined cylinder

Across all cases, the FSI plots reveal clear trends governed by material type and geometric parameters such as L/D ratio and confinement diameter. Increasing the confinement diameter consistently leads to higher fluid velocity magnitudes, as the larger fluid volume allows greater space for momentum transfer and accelerates fluid motion more effectively during collapse. Conversely, increasing the L/D ratio promotes more symmetric and axis-aligned collapse modes, as seen in smoother pressure decays and velocity contours concentrated near the midspan, indicating a slower and more uniform implosion process. Titanium cylinders, due to their higher stiffness, exhibit more intense pressure drops and sharper velocity gradients, reflecting localized collapse and higher-order buckling modes. These velocity patterns also suggest differences in collapse mechanisms: lower L/D and stiffer materials tend to trigger abrupt, lobed buckling and radial jetting, while higher L/D geometries collapse through more global and progressive deformation. These trends highlight that confinement geometry and material stiffness jointly dictate whether the collapse behaves in a localized or distributed manner, which directly impacts shock transmission and energy dissipation. From a design perspective, tailoring cylinder slenderness and confinement dimensions offers a means to modulate collapse intensity, fluid jetting, and stress concentration which are critical factors in developing structures for high-pressure marine environments or implosion-mitigating enclosures.

## 4. CONCLUSIONS

Hydrostatic implosion of thin-walled metallic cylinders in semi-confined underwater environments produces complex fluid–structure interaction phenomena governed by the coupled effects of material stiffness, structural slenderness, and confinement geometry. This study established a validated numerical framework using the structured ALE formulation in LS-DYNA to capture collapse initiation, buckling mode evolution, pressure wave propagation, and energy transfer mechanisms. Direct comparison with controlled experiments demonstrated quantitative agreement in collapse pressure of 3.69 MPa and first water hammer amplitude with an error of 1.01 percent, confirming the predictive fidelity of the model for semi-confined configurations.

The results demonstrate that material properties strongly influence implosion intensity. Titanium cylinders stored greater strain energy prior to collapse and released it more abruptly, generating peak water hammer pressures exceeding 70 MPa in low L/D configurations. Aluminum cylinders exhibited comparatively smoother pressure decay and reduced fluid acceleration. Structural slenderness governed collapse mode transitions and energy release timing, with lower L/D ratios producing sharper kinetic energy peaks and stronger radial jetting, while higher L/D ratios promoted more distributed deformation and delayed energy transfer. Confinement diameter modulated both pressure amplification and fluid mobility, highlighting the non-trivial interaction between geometric restriction and energy redistribution.

Overall, the study provides experimentally validated insight into confinement-amplified implosion dynamics and establishes a reliable predictive framework for assessing collapse-induced shock loading in subsea pressure vessels, marine pipelines, and confined underwater structural systems. The present framework provides an excellent opportunity for future research to conduct systematic comparative investigations of confined, semi-confined, and unconfined implosion scenarios, enabling deeper understanding of the mechanisms governing collapse dynamics, pressure wave propagation, jet formation, and fluid–structure interaction under varying boundary proximity conditions. Furthermore, the numerical simulations presented in this study enable the generation of high-resolution datasets across a diverse range of material and geometric configurations. Unlike experiments, which are often limited by cost, repeatability challenges, and measurement constraints, simulations enable the controlled extraction of detailed pressure, velocity, and energy fields under systematically varied conditions. This capability is particularly advantageous for developing machine learning–based predictive models, which require large, diverse, and physically consistent datasets for robust training. By leveraging simulation outputs, surrogate models can be trained to rapidly estimate key implosion metrics such as peak dynamic pressure, fluid velocity magnitudes, and energy absorption characteristics, dramatically reducing the need for repeated physical testing. This integration of physics-based simulation and data-driven modeling presents an excellent opportunity for future research in predictive design and comprehensive assessment of fluid–structure collapse scenarios.

**Acknowledgements**

This research was conducted at the Multiscale and Multiphysics Mechanics Laboratory at the University of Rhode Island (URI). The authors gratefully acknowledge the use of the Andromeda High Performance Computing Cluster at URI for carrying out the numerical simulations presented in this work.